\documentclass{article}

\usepackage{latexsym}
\usepackage{amsmath}
\usepackage{amsfonts}
\usepackage{amssymb}
\usepackage{amsthm}
\usepackage{mathtools}
\usepackage{txfonts}
\usepackage{mathrsfs}
\usepackage{stmaryrd}
\usepackage{float}
\usepackage{tabularx}
\usepackage{rotating}
\usepackage{graphicx} 
\usepackage{hyperref}
\usepackage{subcaption}
\usepackage{adjustbox}
\usepackage[utf8]{inputenc}

\title{The interdependency structure in the Mexican stock exchange: A network approach}

\author{Erick Trevino Aguilar {erick.trevino@im.unam.mx}
\footnote{Unidad Cuernavaca del Instituto de Matem\'aticas, UNAM}}

\newtheorem{theorem}{Theorem}
\newtheorem{remark}{Remark}

\newcommand{\tr}{\mathbf{tr}}   

\newcommand{\reals}{{\mathbb R}}
\newcommand{\nn}{\mathbf{N}} 

\begin{document}
\maketitle
\begin{abstract}
Our goal in this paper is to study and characterize the interdependency structure of the Mexican Stock Exchange (mainly stocks from BMV) in the period 2000-2019 and  provide visualizations which in a one shot provide a big-picture panorama.  To this end, we estimate correlation/concentration matrices from different models and then compute metrics from network theory including eigencentralities and network modularity.
\end{abstract}

\section{Introduction}
In this paper we investigate the interdependency structure of daily returns in the Mexican stock exchange market. To this end, we build a database of free and publicly available time series of main stocks for the period 2000-2019 and conduct our study in two stages that are then put together to give a unified treatment to our main topic of interest here which is the interdependency structure of daily log-returns in the Mexican stock exchange.\\

In the first stage we focus on the estimation of partial correlations of log returns of daily prices. The reason to focus on partial correlations is the following. Given a collection of Gaussian series $A_1,\ldots, A_n$ a zero partial correlation between  $A_1$ and $A_2$ implies that $A_1$ and $A_2$ are conditionally independent  meaning that $A_1$ and $A_2$ could still be (unconditionally) correlated but only through a third factor adapted to the other series $A_3,\ldots,A_n$. There are of course different methods to estimate a covariance/correlation/concentration matrix and we have selected a estimation based on a specific class of Markovian Random Fields (MRF) which in the statistical literature are well known under the name Gaussian Graphical model  (GGm). The adjective ``graphical'' emphasizes the fact that attached to the probabilistic model there is a graph in which edges expresses conditional dependencies, from which  a very convenient visual representation is obtained.  There are three reasons to work with this model.  First of all, the benefit of the already mentioned visual representation provided by the model.  The second is that we have decided to study the period 2000-2019  in a yearly basis. There is a trade-off to this treatment. On the one hand, short periods of time reduce problems with heavy tails. On the other, the number of stocks in each year is a significant proportion of the available observations. Hence, a \textit{lasso}-regularized estimation is useful in this context which is inbuilt in the estimation of a GGm.  Third, we want an estimation that filters out a ``noisy'' correlation selecting only clear relationships between two series, again this is provided by the \textit{lasso}-regularized estimation.  Loosely speaking, we follow a partial correlations selection approach which conceptually is comparable to a covariance selection approach \cite{Dempster1972}. Once partial correlations matrices have been estimated we provide a list of \textit{stylized facts} from them. Then, taking the graphs constructed from the matrix of partial correlations as its adjacency matrix, we compute degree and eigen centralities and rank different stocks accordingly.\\

In the second stage we estimate correlation matrices of time series (estimated through a Multivariate Dynamic Conditional Correlation GARCH specification). Then, apply  a technique from network-theory based on those correlation matrices: The maximization of a  modularity objective function. This procedure will provide with a partition on the stocks list for a community structure.\\

After this introduction the paper is organized in the following form.  Section \ref{sec:Background} gives some background on the approach of random networks in finance and economics. It also provides details on the data used to feed the models. In Section \ref{sec:GGm} we report on the estimated partial correlation matrices from GGm's. In Section \ref{sec:centralities} we report centralities of networks based on the partial correlation matrices from the previous Section \ref{sec:GGm}. In Section \ref{sec:communitydetection} we report on correlation matrices computed from a multivariate GARCH specification and then maximize a modularity objective function of networks based on  these matrices. This will define groups (communities) of stocks. Section \ref{sec:FinancialDiscussion} concludes the paper with a financial discussion based on the main findings of estimations.
\section{Background} \label{sec:Background}
The classical Markowitz theory of portfolio selection illustrates the correlation matrix relevance for financial decisions.   However, it has been longly known  the nontriviality of correlation estimation from empirical data. Moreover, in contexts where sparse correlation (specially for partial correlation) matrices are expected, it is desirable to have a systematic method to discard  ``non-clear correlations'' and  account for a parsimonious model as motivated by \cite{Dempster1972}. As we mentioned in the introduction, in this paper we choose to apply a GGm for a parsimonious  estimation of concentration/partial-correlation matrices. For the estimation of correlation matrices we apply a multivariate GARCH model.\\

Beyond the estimation problem, it is useful to have tools  that starting from matrices are able to generate metrics providing snapshots of the market from which quick but trustable diagnosis are possible. Situations in which this is desirable include, from the point of view of an investor, the decision of re-balancing a portfolio,  and from the point of view of a regulator, interventions in the market in order to lessen the contagion of a shock in a specific sector.\\

We find those tools in the theory of random networks. Specially in the form of  local metrics (centralities computed from concentration matrices) to classify the interconnectedness of stocks and a global metric (the modularity computed from correlation matrices) to detect \textit{communities} of stocks.\\

The described approach is not new in finance and economics, however for the Mexican stock exchange there are few works in this direction. In the next section we present related literature.  Note however that we do not pretend to give an exhaustive list on this active topic which deserves a survey by its own, but to give a brief panorama on activity for this line of work.
\subsection{Stock markets from GGm, random graphs, and network-theory approaches}
Gaussian graphical models, Random graphs, and Network theory approaches in a financial context is an active research area attracting more and more attention with an increasing number of papers. The following is a non exhaustive list just exhibiting the different approaches and applications. 

Papers with a GGm approach  in finance include  \cite{Giudici2020}, \cite{Denev2019}, \cite{Giudici2016}. Theoretical background on graphical models can be found on \cite{Wainwright2008}, \cite{Lauritzen1996}, \cite{Andersen1995}. Some studies determine power laws for degree connectedness defined by  assets correlation matrices; see \cite{Millington2020}, \cite{Tse2010},  \cite{Jiang2010}, \cite{Boginski2005}, \cite{Onnela2003}, \cite{Bonanno2001}. Papers studying community detection in a financial context include \cite{Millington2020}, \cite{MacMahon2015}, \cite{Almog2015}, \cite{Isogai2014}, \cite{PICCARDI2011}. See \cite{Fortunato2016} for a survey on methods for community detection. Minimal Spanning Trees applied to financial market ranking include \cite{Guo2018}, \cite{Li2018}, \cite{Wang2018}, \cite{Bonanno2001}, \cite{Mantegna1999}. Spillover effects and shocks contagion, \cite{Shahzad2018}, \cite{Narayan2017}, \cite{Kim2015}, \cite{Anufriev2015}, \cite{Dewandaru2014}. 
Portfolio selection, \cite{Peralta2016}, \cite{Pozzi2013}. Detection of stock prices manipulation \cite{Shi2019}. Portfolio diversification \cite{Boginski2005}.
Also relevant for statistical analysis, random matrix theory for  correlation matrices has been studied in \cite{Anufriev2015}, \cite{MacMahon2015}, \cite{Sandoval2012}, \cite{Plerou2002}.

\subsection{Subprime crisis}\label{subs:subprimecrisis}
\begin{figure}[H]
\begin{center}
\includegraphics[width=10cm]{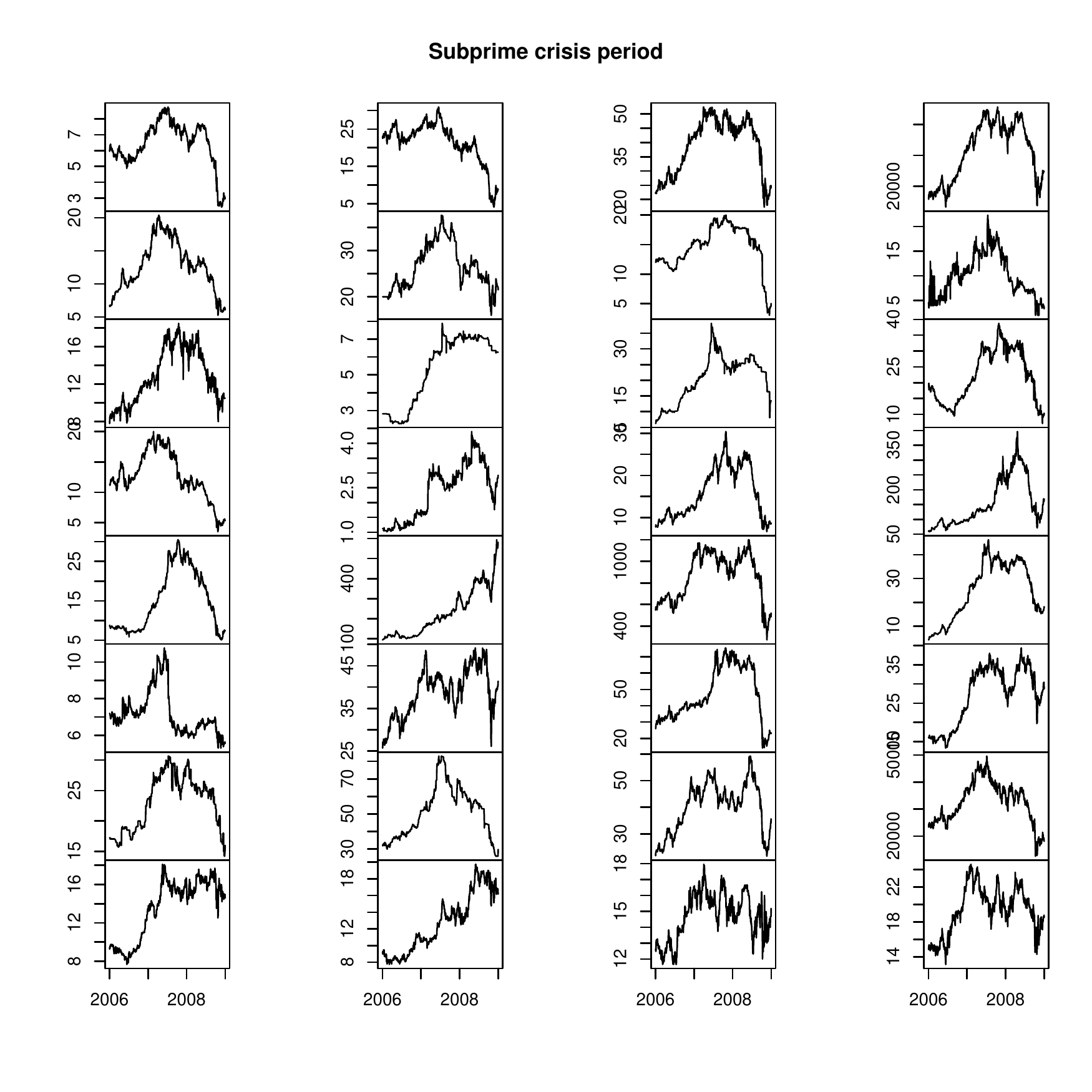}
\end{center}
\caption{Filtered list of stock time series for the period 2006-2008.}
\label{Fig2006}
\end{figure}

According to \cite{Sidaoui2010} there indeed existed an impact from the  2007-2008 subprime crisis in the  Mexican economy mainly due to two shocks, first, a decline in Mexico's exports and second, a constrained access to international financial markets. Thus, an event evidencing an integration of the Mexican stock exchange and the US market. A phenomenon documented by some authors; see e.g., \cite{RomandelaSancha2019,Salgado2017,Vodenska2016}. 
Figure \ref{Fig2006}   illustrates  price levels for some selected stocks in the Mexican stock market for the years 2006, 2007 and 2008. It can be argued that on 2006 bullish stocks dominate the market while on 2008 they turn bearish. Not  surprising and reported by some authors \cite{Perales2017,JaramilloOlivares2016}. 
Later we will go beyond a visual examination and confirm by a multivariate GARCH model through a shift from positive to negative intercepts on log returns of each time series of the period; see Section \ref{sec:trend20062008} below.  However, quite interesting, we will show that the partial-correlations interdependency structure  of the Mexican financial market does not show a dramatic change as response to that shock for price levels, see Figure \ref{FigGGm1} and Section \ref{sec:GGmStylf}.   
\subsection{Data}
We constructed a database of close prices from publicly available information at Yahoo.Finance website. The complete list of analyzed stocks can be found in Appendix \ref{sec:Selectedstocks} while the database and all estimations are themselve available upon request.  The frequency was daily in a span of time comprising 01-01-2000 to 31-12-2019. Always considered time series of log returns: $R_t(i)=\log\left( \frac{S_{t+1}(i)}{S_t(i)}\right)$ where $S(i)$ is the price level of stock labeled $i$.\\
We breakup the data in windows of one year (from january to december) and applied a filtering process in two steps.  In the first step, for each year, stocks in the market  with the most complete information were selected. The criterion was that only stocks with more than 90\% of all the available dates were selected. Then, in  a second step, stocks prices not having a minimum of  variance in moving windows spanning 30 dates were discarded. This filtering process already presents the interesting fact of a positive evolution of the Mexican market for equities in the sense of an increase of activity. Indeed, as we go forward along the years, more and more time series of stocks prices satisfy the filtering process evidencing a grow up in terms of more activity in the market with more variability of prices and more quotes. Visual evidence can be found in Figure \ref{FigGGm1} and Section \ref{sec:Selectedstocks}. An important aspect of this work will be to consider how industrial sectors are interconnected.  Here we consider a list of sectors  obtained from a BMV's classification. These are listed in Table \ref{tab:Industrialsectors}.
\begin{table}[H]
\caption{Industrial sectors}
\label{tab:Industrialsectors}
\centering
\begin{adjustbox}{width=\columnwidth,center}
\begin{tabular}{rrrrr}
\hline
Energy & Industry & The IPC Index & Materials &  Basic consumming\\
\hline
Health & Telecommunications & Financial services & Non basic consumming & Information Technologies\\
\hline
\hline
\end{tabular}
\end{adjustbox}
\end{table}

\subsection{Covariance selection}
Let $\Sigma$ be the covariance matrix  of a random vector $(R_1,\ldots,R_n)$ with multivariate Gaussian distribution. A zero component $\Sigma_{i,j}=0$ expresses marginal independence between $R_i$ and $R_j$.  On the other hand, the inverse matrix $J:=\Sigma^{-1}$, the so-called concentration matrix,  has the property that a zero component $J_{i,j}=0$ expresses conditional independence; see e.g., \cite[Thm. 9.2.1]{Maathuis2019} or for complex  distributions \cite[Thm 7.1 p. 117]{Andersen1995}.  This property of multivariate normal distribution is fundamental in covariance selection; see  \cite{Dempster1972}.\\
Latest developments on covariance selection focus on sparse large dimensions in which the number of variables is large but also there are many variables which are conditionally independent; see \cite{Meinshausen2006}. Thus, for such structures the concentration matrix is sparse and the lasso (least absolute shrinkage and selection operator; also lasso or LASSO) method introduced by \cite{Tibshirani1996}  is fundamental for statistical estimation and variable selection. Indeed, the Gaussian Graphical model that we are going to use is ``nodewise estimated'' through a lasso procedure.  For the lasso implementation we use R  package mgm that builds on the package glmnet. The estimations of this last package are based on the algorithm of \cite{Friedman2007}. Then, the collection of nodewise regressions are combined through an AND rule to give a unique estimation of a multivariate vector. This approach is naturally based on the asymptotic consistency results due to \cite{Meinshausen2006}.   In particular, the estimation yields a concentration matrix $J$. Systematic presentations for graphical models can be found in \cite{Wainwright2008}, \cite{Lauritzen1996}, \cite{Andersen1995}.

\section{GGm}\label{sec:GGm}
\subsection{Markovian Random Fields}\label{sec:GGm:randomFields}
In this section we start with the basic definition of a Markovian Random Field (MRF) which is the fundamental probabilistic concept  from which a GGm is defined.  Let us introduce a graph $G=(V,E)$ with a set of nodes $V=\{1,\ldots,n\}$ and edges $E$. Recall that a complete subgraph of $G$ is called a \textit{clique}.  We denote by $\mathcal C$ the class of \textit{maximal cliques} of the graph $G$.  Let be given a random vector $\vec X=(X_1,\ldots,X_n)$ with multivariate accumulative distribution  function $p$.  Then, the vector $\vec X$ has a \textit{Gibbs distribution} compatible with the graph $G$ if it has a representation
\[
p(x_1,\ldots,x_n) = \frac{1}{Z} \prod_{C  \in \mathcal C} \psi_C(x_C),
\]
where $\{\psi_{c}\}_{C \in \mathcal C}$ are suitable functions and $x_C$ denotes a vector in which only the indexes of $C$ appear. Gibbs distribution are characterized through different Markov properties. Similarly $X_{A}$  for $A \subset V$, $A=(A_{i_1},\ldots,A_{i_k})$ denotes the vector $(X_{i_1},\ldots,X_{i_k})$. Let us introduce them:
\begin{enumerate}
	\item $\vec X$ is a MRF with respect to $G$ if it has the Markov property: For any pair $i,j \in V$  with $i\neq j$ and non adjacent in the graph $G$, the random variables $X_i$ and $X_j$ are conditionally independent on all the other variables. We denote this conditional independency by:
	\[
	X_u \Perp  X_{v} \mid X_{V/\{u,v\}}.
	\]
	\item $\vec X$ is locally a MRF with respect to $G$ if: For each $v \in V$, the random variable $X_v$ is conditionally independent of all other variables which are not neighboors (they are not adjacent). We denote this by
	\[
	X_v \Perp  X_{V/neighborhood(v)} \mid X_{neighborhood(v)}.
	\]
	\item $\vec X$ is globally a MRF if: For two disjoint subsets $A, B \subset V$, the vectors $\vec X_A$, $\vec X_B$ are conditionally independent on a separating set $S \subset V$. We denote this by:
	\[
	X_A \Perp X_B \mid X_S.
	\]
\end{enumerate}
The next is a fundamental equivalence result; see \cite[Chapter 7]{Grimmett2010}.
\begin{theorem}[Hammersley-Clifford]
	Assume that the distribution $p$ of $\vec X$  is defined in a finite state space and is  positive  valued. Then $p$ is a Gibbs distribution if and only $\vec X$ satisfies any of the Markov properties.
\end{theorem}

For a list of Gibbs distributions see e.g., \cite[Section 3]{Wainwright2008}. In this paper we will work with the following specific Gibbs distribution (hence, specific MRF and specific GGm)
\begin{equation}
\label{eq:GMRF}
p_{\theta}(x)=\exp\left\lbrace \theta \cdot x   +   \frac{1}{2}\sum_{i=1}^{m} \sum_{i=1}^{m} \Theta_{i,j} x_i x_j  -A(\theta) \right\rbrace.
\end{equation}
where $A(\cdot)$ is a normalizing constant; see \cite[Example 3.3]{Wainwright2008} for more details. The MRF model in \eqref{eq:GMRF} specifies also the GGm we will work with. Indeed, \eqref{eq:GMRF} does not apriori specify any graph, but from the set of parameters $\Theta_{i,j} \in \reals$ we derive a partial correlation matrix which indeed can be seen as the adjacency matrix of a weighted graph.

\subsection{The GGm}
Now we explain the specification of the GGm we are going to estimate. Let $\Sigma$ be the covariance matrix of the log returns time series $R(1),\ldots,R(n)$. Denote by $J$ the concentration matrix, $J:=\Sigma^{-1}$.  Indeed, the components of the matrix $J$ are given in terms of the coefficients $\Theta_{i,j}$ in equation \eqref{eq:GMRF}. Denote by $\rho_{i,j}$ the partial correlation of $R(i)$ and $R(j)$. Consider the linear regressions defining partial correlations: \begin{equation}\label{eq:RegforParCor}
R(i)-\mu(i)= \sum_{j\neq i} \beta_{i,j}(R(j)-\mu_j) +\epsilon(i)
\end{equation}
where $\mu_i$ is the unconditional mean of $R(i)$ and $\epsilon(i)$ is a residual. Then
\begin{equation}\label{eqbeta}
\beta_{ij}=\frac{\rho_{i,j}}{\sqrt{var(\epsilon(i))var(\epsilon(j))}}.
\end{equation}
It is also true that  
\[
\rho_{i,j}=\frac{-J_{i,j}}{\sqrt{J_{ii} J_{jj}}}.
\]
The adjacency matrix $\mathbf P=(P_{i,j})$ is defined by 
\begin{equation}\label{eq:adjacencymatrix}
P_{i,i}=0 \mbox{ and } P_{i,j} =\rho_{i,j}.
\end{equation}

\begin{remark}
Let us emphasize now that the estimation of the GGm \eqref{eq:GMRF} will ultimately result in the matrix $\mathbf P$ and this matrix is our main input for this section.
\end{remark}
\subsection{Results from GGm estimation: Stylized facts}\label{sec:GGmStylf}
\begin{figure}[H]
\centering
\includegraphics[width=10cm]{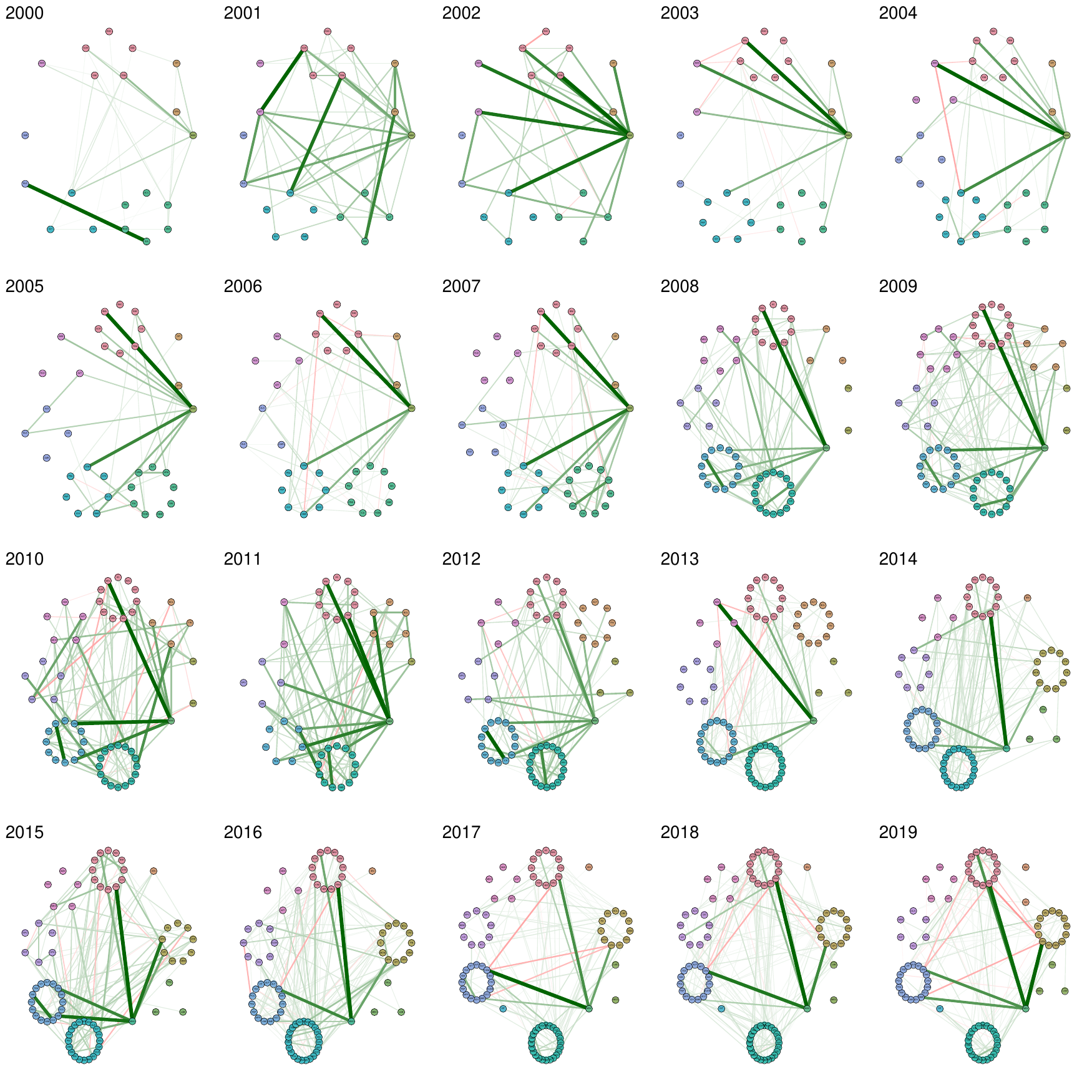}
\caption{Graphs associated to partial correlation matrices by year in the period 2000-2019. For a given edge the green color  (resp. red color ) represents a positive (resp. negative) relationship. Edge's width represents strength of correlation.  A list of partial correlations in different ranks can be found in Appendix \ref{sec:Partialcorrelations}. Vertexes are grouped according to its industrial sector.} 
\label{FigGGm1}
\end{figure}

\begin{figure}[ht]
\centering
\includegraphics[width=7cm]{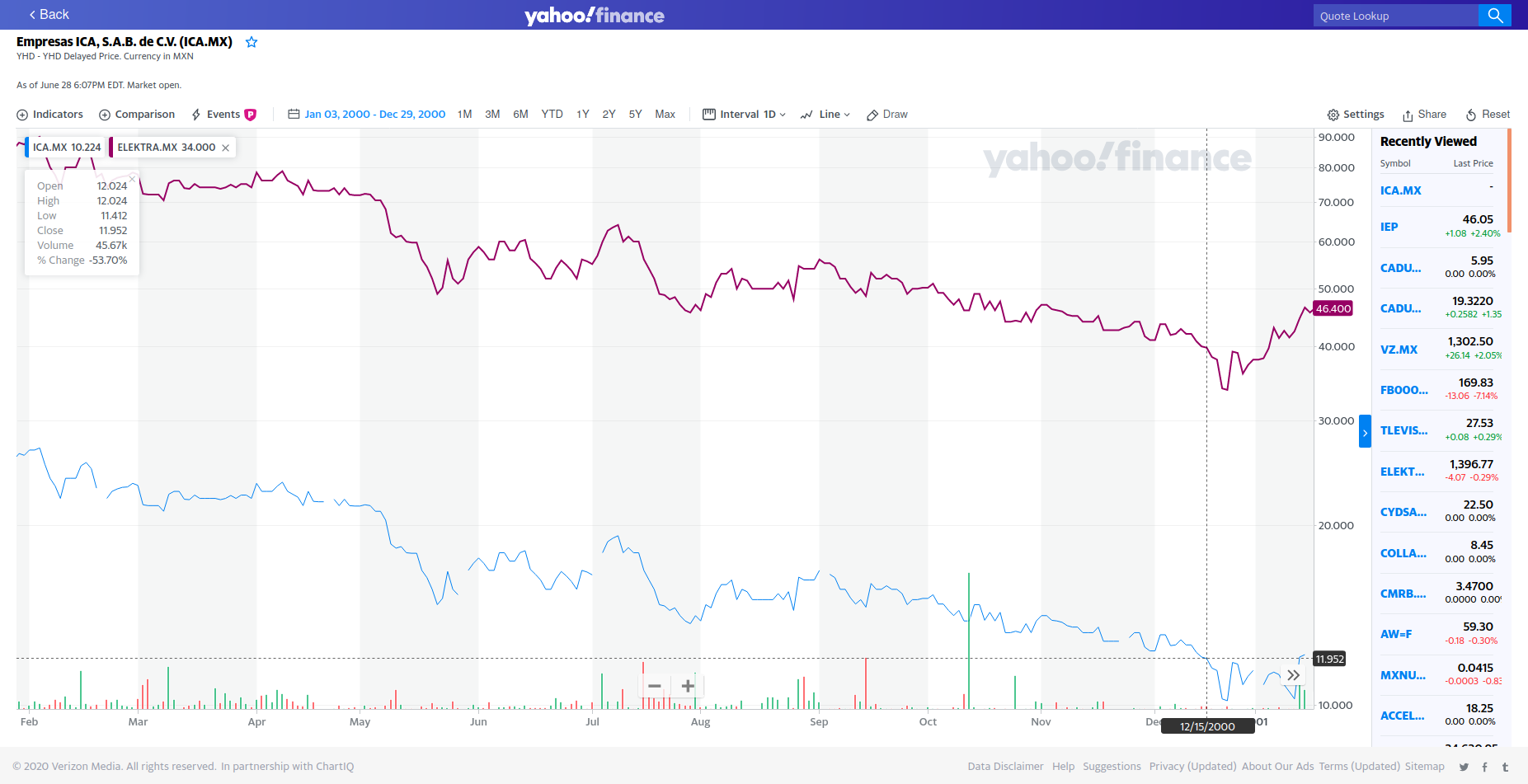}
\caption{In purple the time series for ELEKTRA and in blue the time series for ICA.  Prices in logarithmic scale for the year 2000. Source: Yahoo.Finance.} 
\label{FigGGmicaVSelektra}
\end{figure}

In this section we report the results on estimating a  GGm with underlying MRF (for each year in the period 2000-2019) with specification \eqref{eq:GMRF}. From this estimation exercise, proposed as able of capturing conditional (in)dependencies for logreturns time series, we get a list of partial correlation matrices $\mathbf P$ for the  years in the period 2000-2019.  Such matrices are available as supplementary material. A graphical representation of partial correlations is displayed in the panel of Figure \ref{FigGGm1} and a complete list of most strong partial correlations in the  interval $[0.3,1]$ and $[0.2,0.3]$ can be found in Appendix \ref{sec:Partialcorrelations}.  The complete list is available as supplementary material. From them, we have the following stylized facts:
\begin{itemize}
\item First of all we see in Figure \ref{FigGGm1} a stable continuous evolution of partial-correlations interdependence structure.   In particular, at this step of a visual evidence, if there indeed exist impact of global crisis episodes (e.g., dot.com bubble, Subprime crisis, etc.) it doesn't seem to affect at large for the partial-correlations interdependence structure.

\item Many links above 0.2 and frequently include the  main index from the Mexican stock Exchange BMV denominated IPC (quoted as MXX in Yahoo.Finance);  see the tables in Appendix \ref{sec:Partialcorrelations}.

\item As we move forward in time, the market grows (with more nodes of stocks consistently quoted by year). However, it does not seem to be evidence that interconnectedness in the market changes drastically from one year to the other, even for the subprime crisis period. 
 
\item Stock connections must be due to exogenous factors to the market, since the graph is based on partial correlations.  However, for links that involve as a node the IPC, it typically happens that the other node is a stock involved in the construction of the index.

\item A large number of links between stocks in different sectors. An empirical fact reported for other markets; see e.g., \cite{Millington2020}. To our best knowledge, not previously documented for the Mexican market. Nonetheless, satisfactorily, intrasectorial partial correlations are also present.

\item We see persistent links between pair of stocks that along the twenty years period appear frequently but not systematically; see Apendix \ref{appendix:Endurablelinks}.
\item Negative (partial) correlations appear only seldom.

\item For the year 2000 we see a partial correlation  of 0.98 between ICA and ELEKTRA which apriori looks as an odd finding. But this is actually supported by data; see Figure \ref{FigGGmicaVSelektra}.  

\item The strongest links above 0.3 are those typically having as one of its nodes the IPC. Also for the rank $[0.2,0.3]$ links with IPC dominate but with a little decline in comparison with the interval $[0.3,1]$.

\item For the rank $[0.05,0.2]$, we count  646 links in the period 2000-2019.  The  IPC links are quite rare, only ten appearances.  

\item FEMSA indeed has a persistent relationship with IPC with partial correlations above 0.3; see  the table in Appendix \ref{appendix:Endurablelinks}. In this same table we do not see an important stock as AMX. This is an interesting confirmation for GGm model' strength,  since it captures a realistic fact; see e.g., the news stories \href{https://expansion.mx/empresas/2016/08/24/quien-manda-en-el-principal-indice-de-la-bmv}{expansion} and \href{https://www.eleconomista.com.mx/mercados/Femsa-la-empresa-con-mayor-peso-en-el-IPC-20190509-0114.html}{el economista}, etc.
\end{itemize}

In the tables from Appendix \ref{appendix:Endurablelinks} we see the most ``persistent'' relationships between stocks for which partial correlations in absolute value were in a given interval for nine or more years. Quite notoriously they are rare and always involves the index IPC. We are particularly interested in obtaining metrics  (centralities) from network theory to see a possible effect of the afore mentioned global episodes. 
\section{Network theory: Centralities}\label{sec:centralities}
Centrality is a measure conceptually designed in such a way that a vertex with high centrality is arguably highly influential.  The first concept of centrality we use  is the \textit{degree centrality} which for a vertex in a weighted network is just the sum of all connecting edge's weights.  For our graphs of partial correlations, the degree centrality  gives information of the pattern of a shock's transmission. The idea is that for an influential (i.e., with high centrality) stock in the financial network, a bad day is accompanied with many other stocks in the same situation. Note that there is no causation claimed here. The second  measure of centrality that we estimate is the \textit{eigencentrality}. This is a global measure in that scores for each node are assigned by a contrast of the quality of its links. For example a node with just one link to another influential node could have a highest eigencentrality  than a node with two or more links. The computation of eigencentralities  transfers to a spectral analysis of the adjacency matrix and in crucial steps is substantiated by Perron-Frobenius theory (see e.g., \cite[Chapter 17]{Shapiro2015}). 

\subsection{Shock transmissions}
Let us explain with more detail about eigencentrality and at the same time also clarify about shocks transmission. Let $V=\{1,\ldots,n\}$ denote our set of stocks and recall the matrix $\mathbf P$ defined in \eqref{eq:adjacencymatrix}. The eigencentrality is a function $f:V \to \reals$ satisfying 
\begin{equation}
\label{eq:ideaEC}
f(v)=r \sum_{w \in N(v)} \mathbf P_{v,w}f(w),  v\in V,
\end{equation} 
where $r$ is a non negative constant and $N(v)$ denotes the neighbors of $v$.  Note that 
\[
f(v)=r \sum_{w \in V} \mathbf P_{v,w}f(w),
\]
since by definition $w \in N(v)$ if and only if $\mathbf P_{v,w} \neq 0$. Now this can be written in matricial notation as
\[
f(V)=r \mathbf P f(V)^{T},
\]
where $f(V)=(f(1),\ldots,f(n))$. Hence, $f(V)$ is an eigenvector of $\mathbf P$ attached to $r$ as its eigenvalue.\\

To continue we follow the discussion in \cite{Anufriev2015}, recall the coefficients $\beta_{i,j}$ in equation \eqref{eqbeta}. The matrix of coefficients $\mathbf B=(\beta_{i,j})$ with $\beta_{ii}=0$ is then connected to the adjacency matrix as $\mathbf B=diag(J)^{-\frac{1}{2}} \mathbf P diag(J)^{\frac{1}{2}}$. We can write the linear regression in a compact matricial notation as 
\begin{equation}\label{eq:matricialnotationBJ}
R-\mu=\mathbf B (R-\mu) + \epsilon=diag(J)^{-\frac{1}{2}} \mathbf P diag(J)^{\frac{1}{2}} (R-\mu) + \epsilon.
\end{equation}

Let $\tilde R(i):=R(i)-\mu(i)$ and $\tilde R=(R(1)-\mu(1),\ldots,R(n)-\mu(n))$. Then,
\[
diag(J)^{\frac{1}{2}} \tilde R=  \mathbf P diag(J)^{\frac{1}{2}} (R-\mu) +  diag(J)^{\frac{1}{2}} \epsilon.
\]
Hence the vector $X:= diag(J)^{\frac{1}{2}} \tilde R$ satisfies
\[
X = \mathbf P X + \eta
\]
where $\eta:=diag(J)^{\frac{1}{2}} \epsilon$.\\

Now assume that between times $t_0$ and $t_1$ there is a shock $\Delta=(0, \ldots,0,\delta,0,\ldots,0)$ affecting $X(i)$. Then,  $X$ at time $t_1$ is given by  $\mathbf P( X + \Delta) + \eta$ and the change is then $\mathbf P \Delta$. Note that  $\mathbf P \Delta$ does not need to be a scalar of $\Delta$, meaning that the shock affecting originally to $X(i)$ is also affecting to other components indicating that the shock propagates.  

The spectral decomposition of $\mathbf P$ helps on assessing the reach of propagation and rationalizes the definition of eigencentrality.  Let $W_1,\ldots, W_n$ denote the set of eigenvectors of $\mathbf P$ and $\Lambda=\{\lambda_1,\ldots,\lambda_n\}$ the corresponding set of eigenvalues, which we assume is decreasingly ordered by its modulus. Here is a main assumption: there is a unique eigenvalue attaining the spectral radius. This means  $|\lambda_1|> |\lambda_2| \geq |\lambda_2| \ldots  \geq |\lambda_n|$. If the matrix $\mathbf P$ has only nonnegative components, Perron-Frobenious theory guarantees we are in this situation and even more properties; see e.g., \cite[Chapter 17]{Shapiro2015}. Represent $\Delta$ by $\Delta =\sum_{i} \alpha_i W_i$. Then, for $k \in \nn$
\[
\mathbf P ^k \Delta= \lambda^k_1 \left\lbrace \alpha_1 W_1 + \sum_{i=2}^{n}  \left( \frac{\lambda_i}{\lambda_1}\right)  ^k W_i\right\rbrace.
\]
Hence
\[
\lim_{k \to \infty} \frac{1}{\lambda^k_1} \mathbf P^k \Delta  =\alpha_1 W_1.
\]
Then, as time runs the leading term indicating the effect of the initial shock $\Delta$ takes the form $\lambda_1^k \alpha_1 W_1$.

\subsection{Results of estimation}
\begin{figure}[H]
\centering
\includegraphics[width=.5\textwidth]{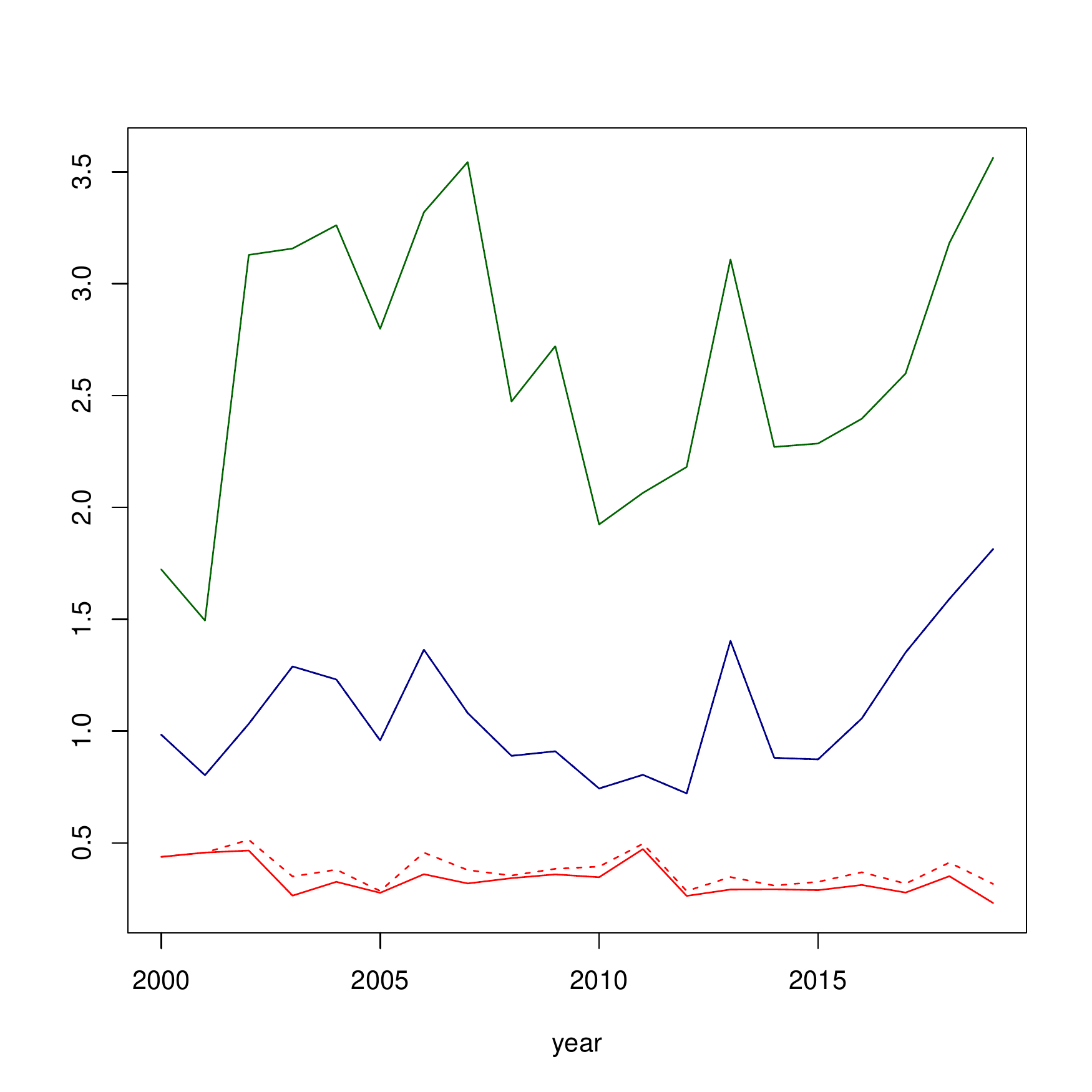}
\caption[first caption.]{The blue line is the largest eigenvalue and dashed blue line is its absolute value. The green line is the maximum degree centrality by year, always attained by the IPC. The red lines are averages of degree centralities, resp. absolute value of degree centralities.}
\label{40centralities1}
\end{figure}

In Figure \ref{40centralities1} we see estimated centralities for our networks.  The  blue line  is the largest modulus per year of eigenvalues. The green line  represents the maximum degree centrality for each year. Very unsurprising this maximum is always  attained by the IPC. The red (resp. red and dashed) line represents the average of each node's degree centrality (resp. average of each node's absolute value degree centrality). 

\begin{figure}[H]
	\centering
	\includegraphics[width=9cm]{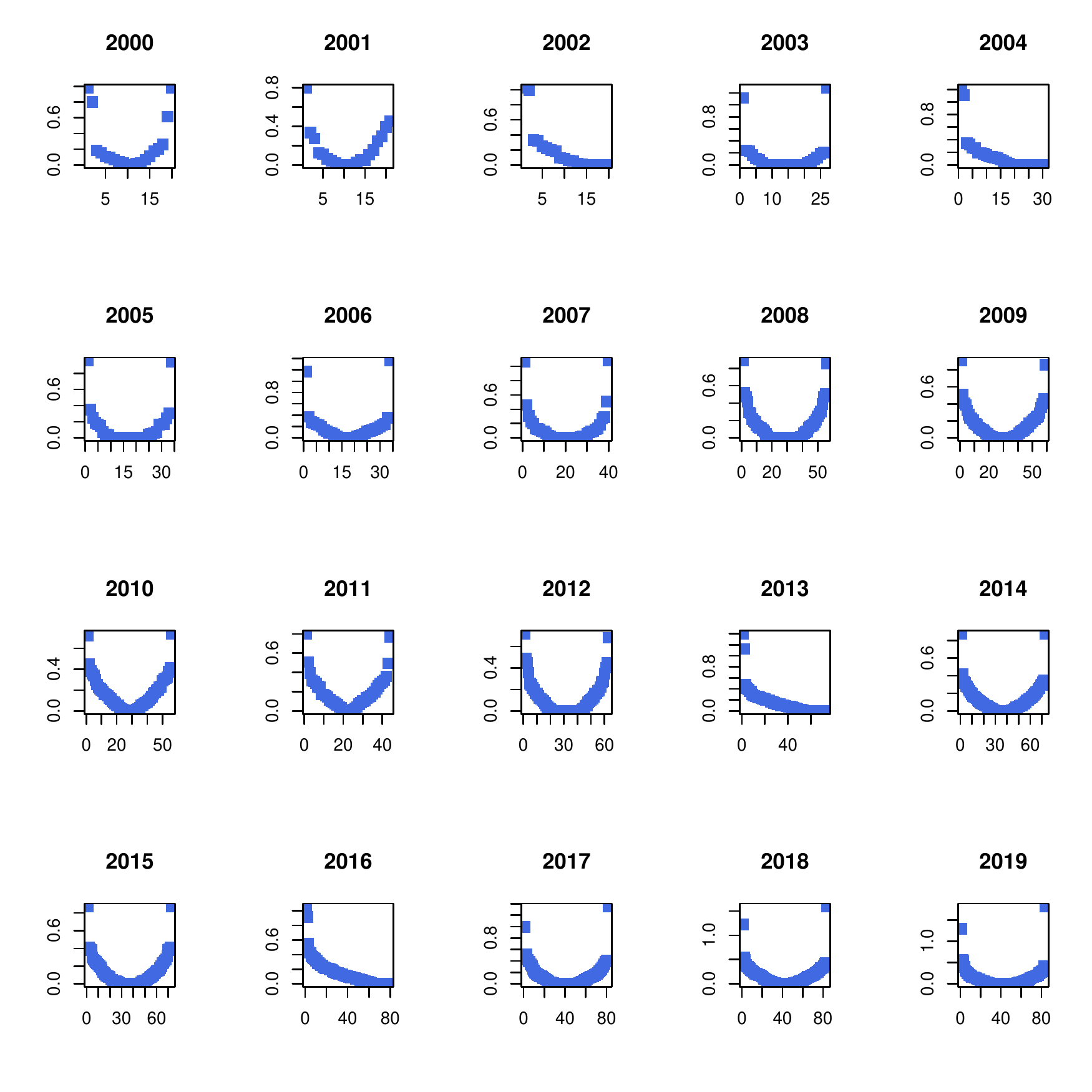}
	\caption{Eigenvalue's modulus per year.} 
	\label{Figplotseigenvalues}
\end{figure}

These are the facts we observe from Figure \ref{40centralities1}:
\begin{itemize}
\item  First of all, spectral radius are approximately bounded by two, which coincides with the range documented for other markets; see e.g., \cite{Millington2020}. 
\item The red line and the red dashed lined are almost indistinguishable. This happens as a consequence to the fact that almost all partial correlations are non negative. We also observe the extraordinary stability on the metric represented by this line.
\item The patterns of the green line and  blue line are similar.  As we mentioned, the green line is attained by the IPC. So it could be expected that also the blue line is related to this index.  Although we do not go into this claim, assuming it is correct, in order to capture effects beyond the IPC it might be necessary in this case to complement with the second eigenvalue together with its eigenvector for centrality and the analysis for a shock contagion. Indeed,  Figure \ref{Figplotseigenvalues} shows that in many cases the dominant eigenvalue has multiplicity two or more, and in other cases that  the second eigenvalue turns out to be close to the first. Certainly, the idea of considering beyond the dominant eigenvector for eigencentrality is not new; see e.g., \cite{Newman2010}. Analysis for the Mexican case will be addressed elsewhere.
\item There is indeed  variability for centralities, but changes from one year to the other, are indeed in units. Thus, changes are subtle. For example, for the subprime crisis period, we see in the blue line small upwards jumps from 2005 ( 0.95) to  2006 (1.36) and then a decline, from 2007 (1.08) to 2008 (0.88). Analogously for max degree centrality in the green line: we see small upwards jumps from 2005 (2.79) to  2006 (3.31) and then a decline, from  to 2007 (3.54) to 2008 (2.47).
\item Continuing with the previous point. We see an abrupt upwards movement for the green line which is reasonable to associate with the dot.com bubble's burst: From the year 2001 (1.49) to 2002 (3.12).
\end{itemize}

\begin{figure}[H]
\begin{minipage}{.5\textwidth}
 \centering
\includegraphics[width=1\textwidth]{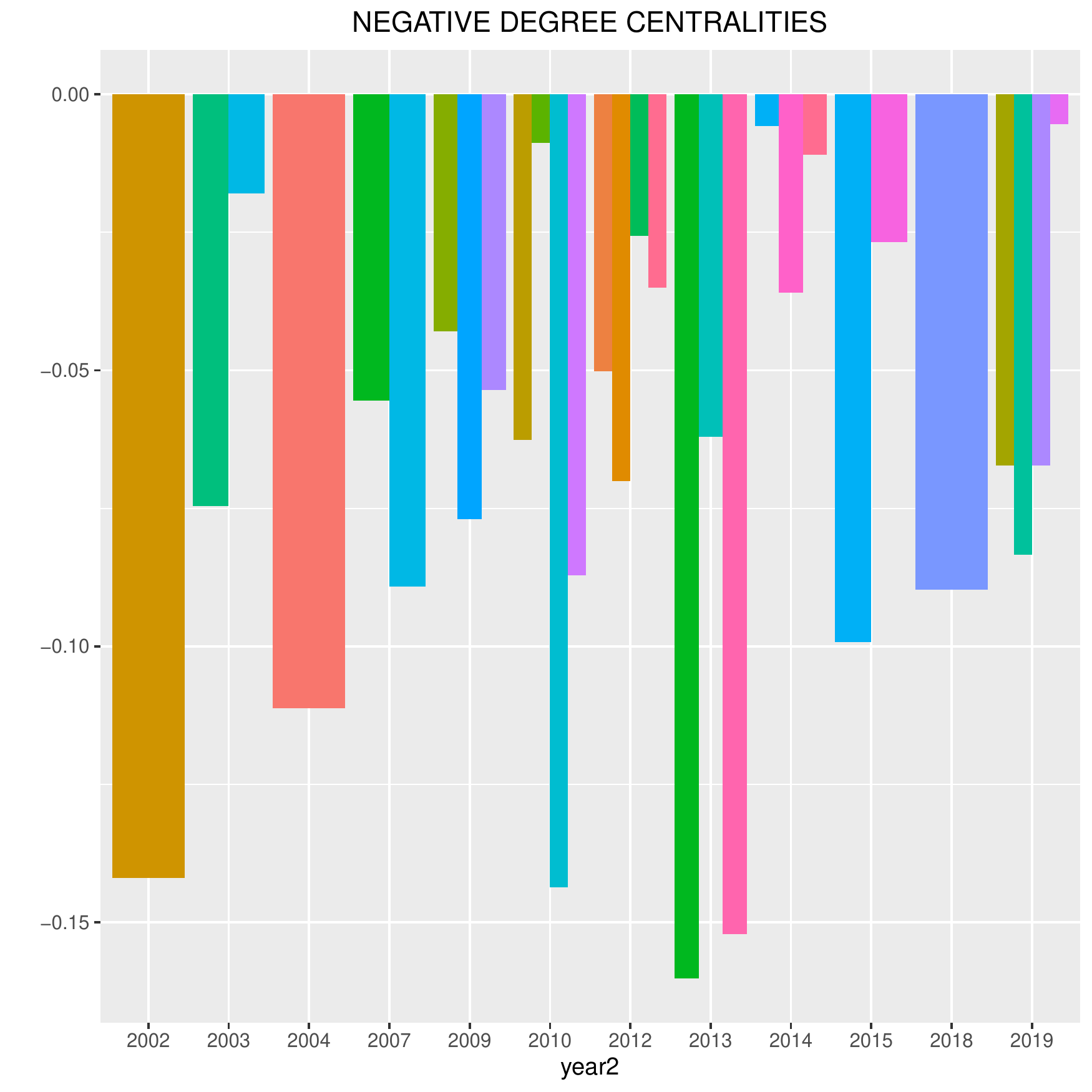}
\subcaption[second caption.]{Only negative values are plotted.}\label{40BarplotnegativeDegreeCentrality}
\end{minipage} 
\begin{minipage}{.5\textwidth}
 \centering
\includegraphics[width=1\textwidth]{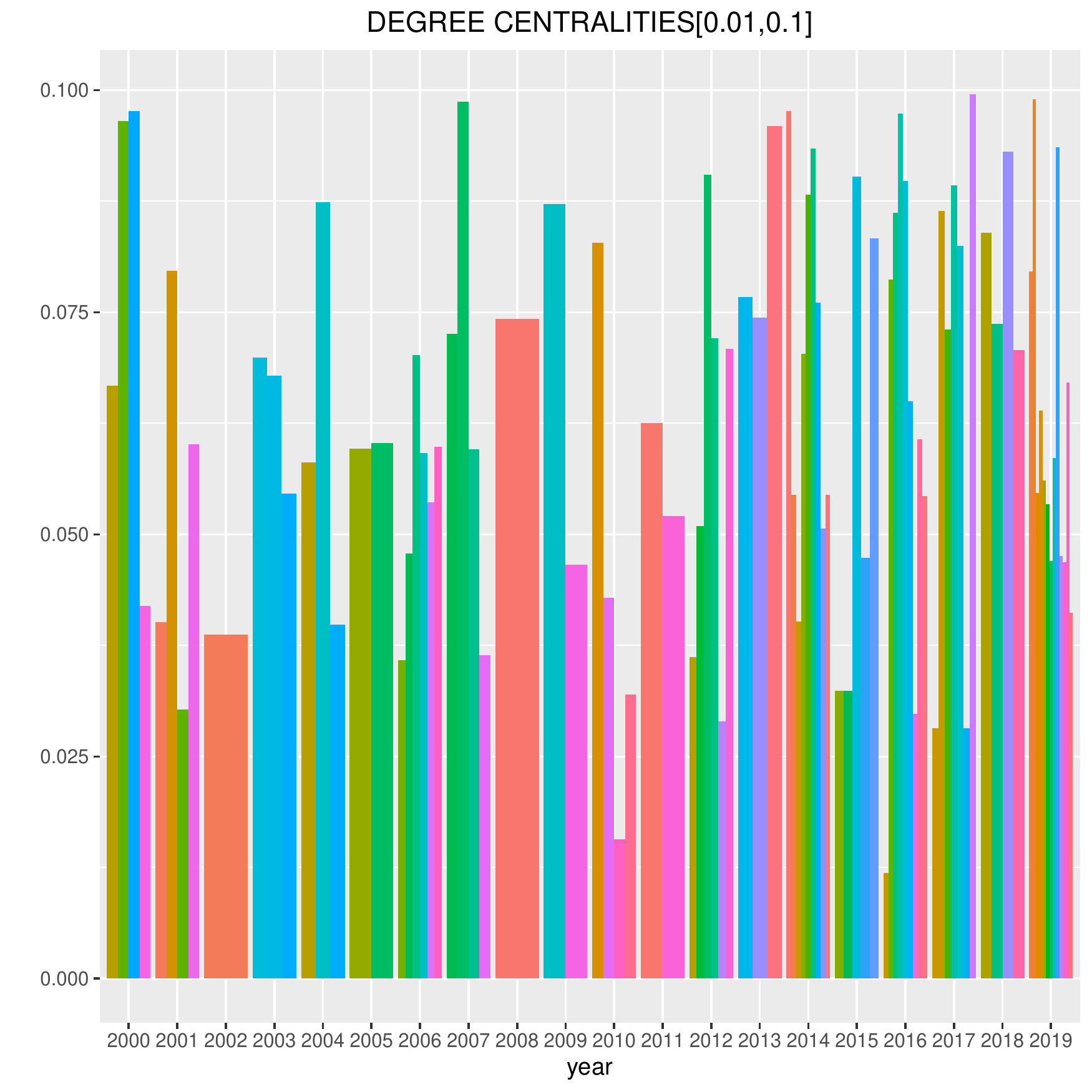}
\subcaption[second caption.]{Values in the range $[0.01,0.1]$ are plotted.}\label{40BarplotpuntocerounopuntounoDegreeCentrality}
\end{minipage}
\begin{minipage}{.5\textwidth}
 \centering
\includegraphics[width=1\textwidth]{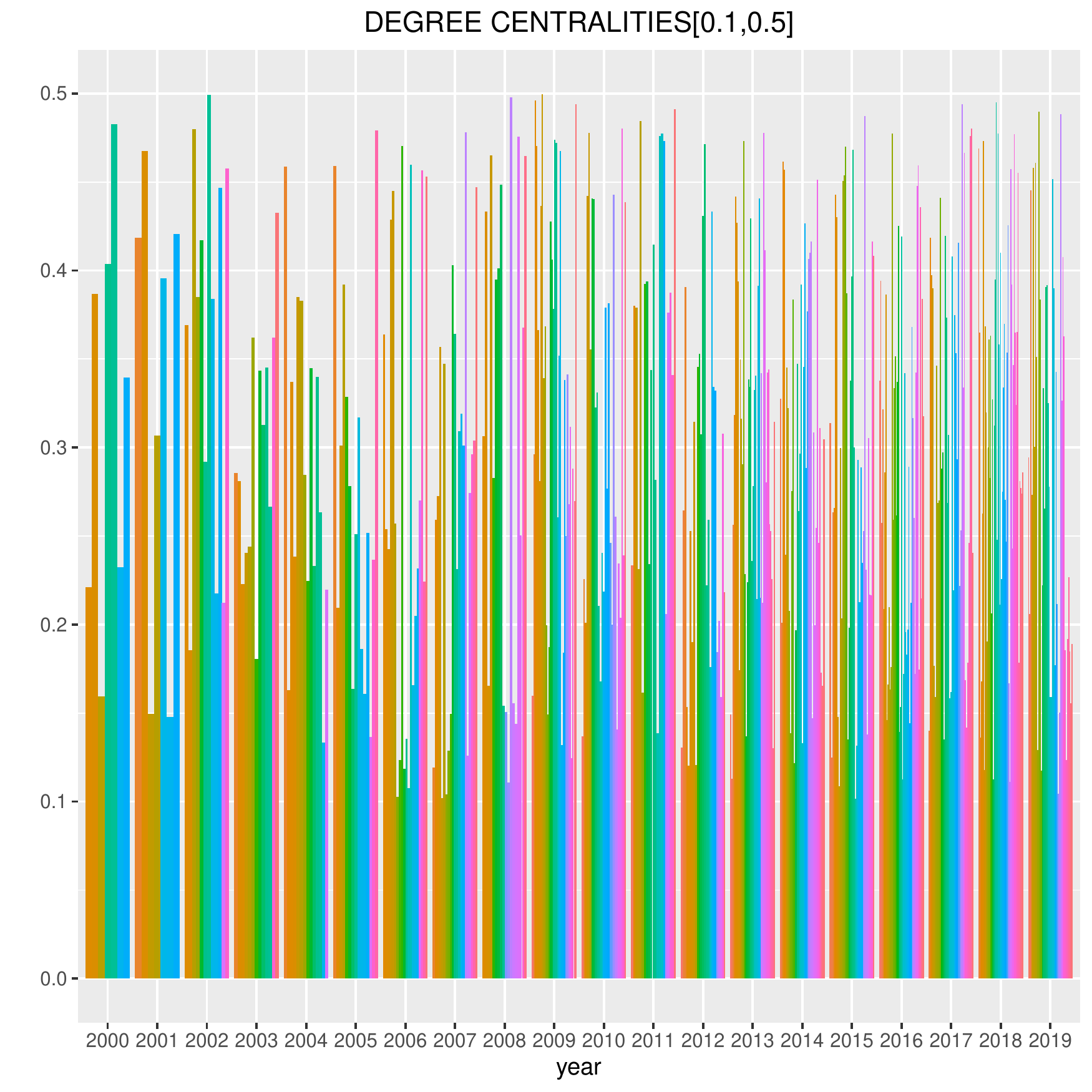}
\subcaption[second caption.]{Values in the range $[0.1,0.5]$ are plotted.}\label{40BarplotpuntounopuntocincoDegreeCentrality}
\end{minipage}%
\begin{minipage}{.5\textwidth}
 \centering
\includegraphics[width=1\textwidth]{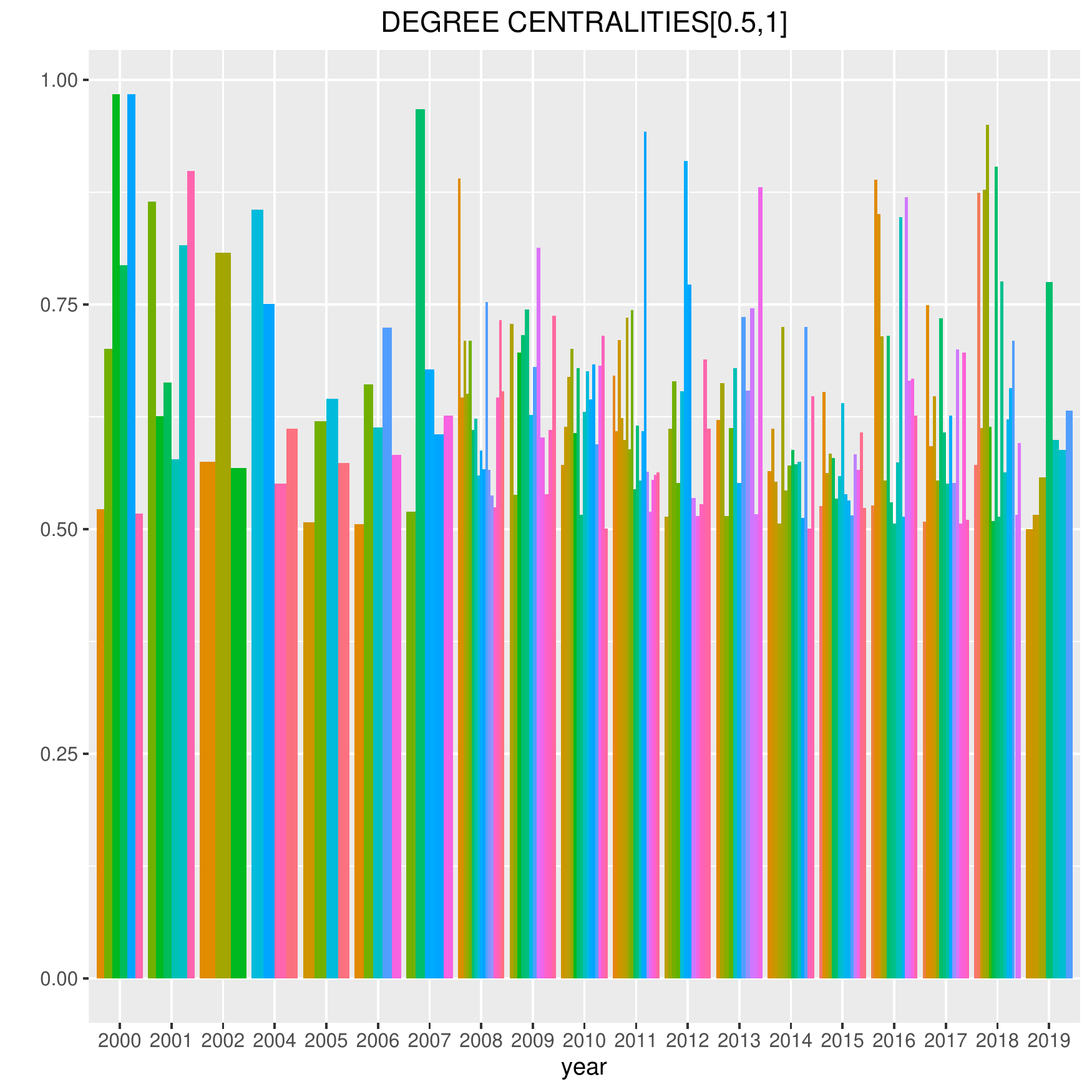}
\subcaption[second caption.]{Values in the range $[0.5,1]$ are plotted.}\label{40BarplotpuntocincoAunoDegreeCentrality}
\end{minipage}%
\caption{A comparison of degree centralities by year at different ranges.}\label{figCentralities}
\end{figure}

\begin{figure}[H]
\centering
\includegraphics[width=8cm]{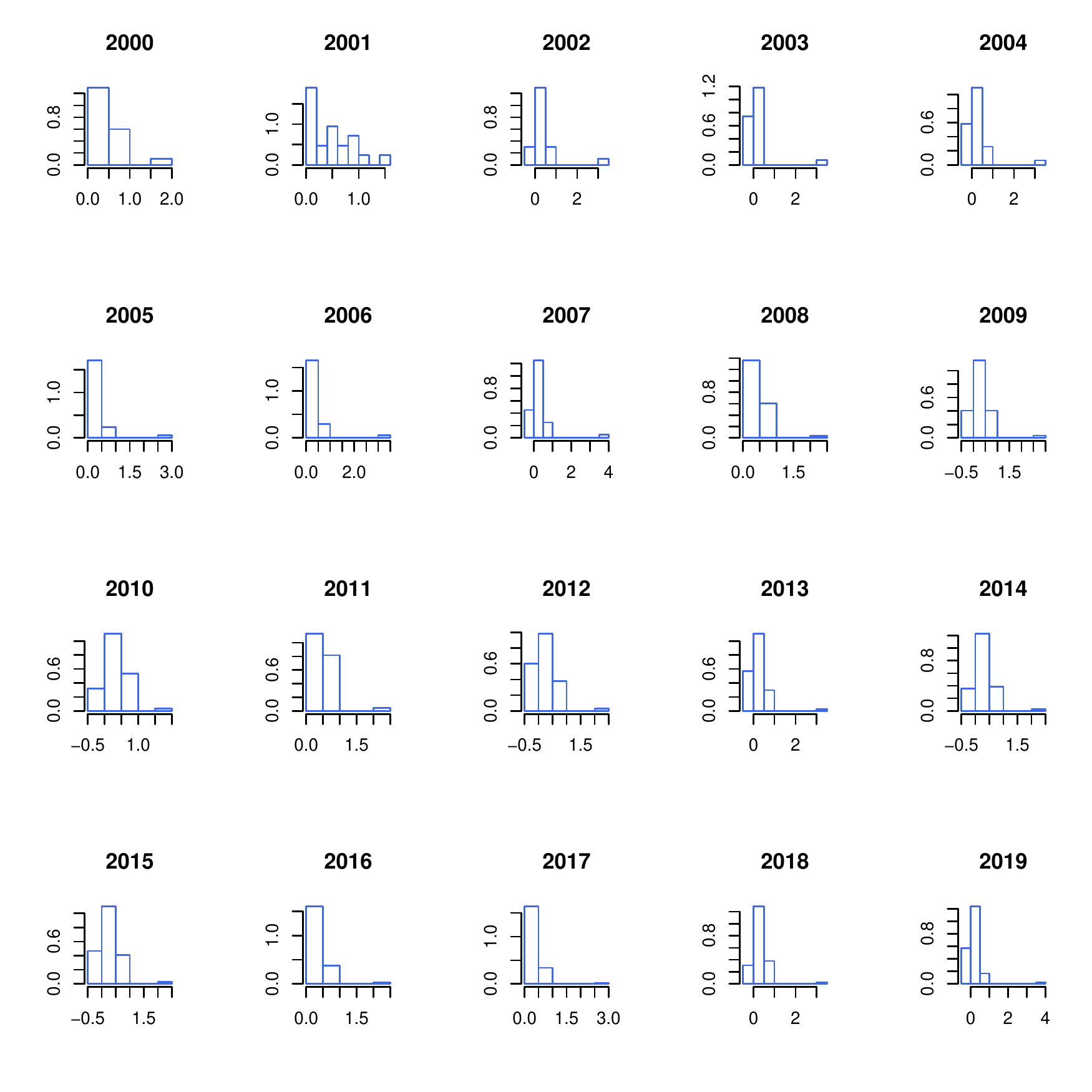}
\caption{Histograms of degree centrality per year.} 
\label{Fig40HistogramDegreecentrality}
\end{figure}

In Figure \ref{figCentralities} we see a panel of barplots for degree centralities separated into different ranges for all stocks in their respective period and in Figure \ref{Fig40HistogramDegreecentrality} histograms per year.  This is what we would like to remark. First of all, as we already mentioned for the red lines in Figure \ref{40centralities1}, links with negative values are few in quantity and magnitude as more precisely illustrated in  Figure \ref{40BarplotnegativeDegreeCentrality}. This is also evidenced in Figure \ref{Fig40HistogramDegreecentrality} where histograms for all years in the period 2000-2019 are illustrated.  In Figure \ref{40BarplotpuntocerounopuntounoDegreeCentrality} we see a quite homogenous distribution in the range $[0.01,0.1]$.  An analogous situation is appreciated in Figure \ref{40BarplotpuntounopuntocincoDegreeCentrality}  in the interval  $[0.1,0.5]$.   Only in the range  $[0.5,1]$ we see in Figure \ref{40BarplotpuntocincoAunoDegreeCentrality} a more heterogenous situation with some dominating  stocks.

\section{Network theory: Community detection}\label{sec:communitydetection}

\subsection{DCC Multivariate Garch model}\label{sec:mGarch}
Let $y:=\{y_t\}_{t=1}^{N}$ denote a one dimensional time series with $N$ observations. A GARCH specification for its volatility usually starts with a flux of information determined by a filtration $\{\mathcal{F}_t\}_{t=1}^{N}$ in which $\mathcal{F}_t$ is a $\sigma$-algebra representing information at time $t$ and $y$ follows the dynamic
\[
y_t=E[y_t \mid \mathcal{F}_{t-1} ] + \epsilon_t(\theta).
\]
Here $\theta$ is a parameter vector whose specification specializes the model, $\mu(\theta)$ is the conditional mean of the time series at time $t$, usually modeled through an ARMA time series. For example an ARMA(1,1) (as we will consider here) is specified by 
\begin{equation}\label{eq:arma11}
\mu_t= \mu + \varepsilon_t + \phi \mu_{t-1} + \psi\varepsilon_{t-1},
\end{equation} 
where $\phi$, $\psi$ are parameters to be estimated and $\varepsilon$ is white noise, i.e. an uncorrelated centered time series. The residual $\epsilon(\theta)$ captures the conditional volatility of $y$:
\begin{align*}
var(y_t \mid \mathcal{F}_t) = E[(y_t - \mu_t(\theta))^2 \mid \mathcal{F}_t]
=E[(\epsilon_t(\theta))^2 \mid \mathcal{F}_t]=var(\epsilon_t(\theta)).
\end{align*}
Its specification is the essence of a GARCH model. We will consider the standard GARCH(1,1) model:
\begin{align}
\epsilon_t  &= \sigma_t z_t\\
\sigma^2_t  &=\alpha_0 + \alpha_1 \epsilon^2_{t-1} +  \beta_1 \sigma^2_{t-1},
\end{align}
where $\{z_t\}_{t=1}^{N}$ is white noise.\\ 

Now consider a set of univariate time series $y(1)$,\ldots,$y(n)$.  A class of models in the \textit{multivariate GARCH} literature known as Dynamic Conditional Correlation (DCC) was introduced by  \cite{Engle2002} and \cite{Tse2002}. The DCC class builds up on  univariate GARCH models and then specifies the dynamic of time varying conditional covariance matrix of the time series $y(1)$,\ldots,$y(n)$. It has a general dynamics
\[
H_t =D_t R_t D_t.
\]
Here $D_t$ is a diagonal matrix of time varying standard deviations from univariate GARCH models and $R_t$ is a time varying correlation matrix.  For estimation, the matrix $R_t$ is decomposed as
\[
R_t= (Q^*_t)^{-1}Q_t(Q^*_t)^{-1}
\]
where $Q$ is specified in \cite[Equation (2)]{Engle2001}.
\subsection{Means for the years 2006 and 2008}\label{sec:trend20062008}
In Table \ref{table2006} we report the coefficient $\mu$ in the specification \eqref{eq:arma11} for each stock in the year 2006, analogously for Table \ref{table2008} in the year 2008. The estimation of these coefficients provides further support to the claim reported  in Section \ref{subs:subprimecrisis} after the visual evidence of Figure \ref{Fig2006}.

\begin{table}[ht]
\caption{The coefficient $\mu$ for the year 2006.} 		
\label{table2006}
	\centering
	\begin{tabular}{rlrlr}
		\hline
		& Stock & mu value & Stock & mu value \\ 
		\hline
		1 & ALFAA & 0.0006 & GISSAA & 0.0011 \\ 
		2 & ALSEA & 0.0026 & GMD & 0.0036 \\ 
		3 & AMXA & 0.0017 & GMEXICOB & 0.0021 \\ 
		4 & ARA & 0.0027 & HERDEZ & 0.0016 \\ 
		5 & AXTELCPO & 0.0013 & HOMEX & 0.0027 \\ 
		6 & AZTECACPO & 0.0006 & ICA & 0.0026 \\ 
		7 & BACHOCOB & 0.0012 & ICHB & 0.0036 \\ 
		8 & BIMBOA & 0.0018 & KIMBERA & 0.0011 \\ 
		9 & CEMEXCPO & 0.0013 & MXX & 0.0021 \\ 
		10 & CMOCTEZ & 0.0016 & PAPPEL & 0.0018 \\ 
		11 & CMRB & 0.0013 & PASAB & -0.0015 \\ 
		12 & CYDSASAA & 0.0016 & PE\&OLES & 0.0027 \\ 
		13 & ELEKTRA & 0.0017 & PINFRA & 0.0065 \\ 
		14 & FEMSAUBD & 0.0024 & RCENTROA & 0.0039 \\ 
		15 & GCC & 0.0022 & SORIANAB & 0.0021 \\ 
		16 & GFINBURO & 0.0007 & URBI & 0.0018 \\ 
		17 & GFNORTEO & 0.0033 & WALMEX & 0.0023 \\ 
		\hline
	\end{tabular}
\end{table}

\begin{table}[ht]
\caption{The coefficient $\mu$ for 2008 year.} 	
\label{table2008}
	\centering
	\begin{tabular}{rlrlrlr}
		\hline
		& Stock & mu value & Stock & mu value & Stock & mu value \\ 
		\hline
		1 & AC & -0.0013 & ELEKTRA & 0.0006 & IDEALB-1 & -0.0008 \\ 
		2 & ALFAA & -0.0019 & FEMSAUBD & 0.0019 & KIMBERA & 0.0002 \\ 
		3 & ALSEA & -0.0013 & FINDEP & -0.0036 & LAMOSA & -0.0016 \\ 
		4 & AMXA & -0.0023 & FRAGUAB & 0.0004 & MAXCOMA & -0.0028 \\ 
		5 & ARA & -0.0018 & GAPB & -0.0017 & MEDICAB & -0.0001 \\ 
		6 & ASURB & -0.0014 & GCARSOA1 & -0.0003 & MEGACPO & -0.0028 \\ 
		7 & AUTLANB & 0.0037 & GCC & -0.0032 & MXX & -0.0010 \\ 
		8 & AXTELCPO & -0.0053 & GFAMSAA & -0.0023 & OMAB & -0.0024 \\ 
		9 & AZTECACPO & 0.0000 & GFINBURO & 0.0008 & PAPPEL & -0.0047 \\ 
		10 & BACHOCOB & -0.0020 & GFNORTEO & -0.0003 & PASAB & -0.0030 \\ 
		11 & BIMBOA & 0.0002 & GIGANTE & -0.0026 & PE\&OLES & -0.0010 \\ 
		12 & CABLECPO & 0.0000 & GISSAA & -0.0009 & PINFRA & -0.0016 \\ 
		13 & CEMEXCPO & -0.0030 & GMD & -0.0053 & POCHTECB & -0.0048 \\ 
		14 & CIEB & -0.0023 & GMEXICOB & -0.0032 & SAREB & -0.0033 \\ 
		15 & CMOCTEZ & -0.0005 & GRUMAB & -0.0001 & SIMECB & 0.0007 \\ 
		16 & CMRB & -0.0005 & HOMEX & 0.0007 & SORIANAB & 0.0010 \\ 
		17 & CULTIBAB & -0.0001 & ICA & -0.0006 & TMMA & -0.0040 \\ 
		18 & CYDSASAA & -0.0029 & ICHB & 0.0010 & URBI & -0.0018 \\ 
		\hline
	\end{tabular}
\end{table}

\subsection{Modularity}\label{sec:modularity}

Assume we are given an undirected and unweighted graph $G$ with vertexes $V=\{1,\ldots,n\}$ and edges $E$. Community structure in the graph means that there exists a partition of $V$ in groups of vertexes in such a way that within groups vertexes are highly connected and more edges exists among them, while at the same time, edges between groups are less observed; see  \cite{Fortunato2016} for a survey of methods in community detection. The afore description presents a general idea and to make it operative, it is necessary to give a more quantitative formulation.  A popular approach is through the famous concept of modularity as introduced by \cite{Newman2004} and further developed in \cite{Newman2006}.  Following the notation of \cite{Newman2006} we introduce the following objects. Let $A$ be the adjacency matrix of $G$ and let $m=\frac{1}{2} \sum_{i} k_i$ where $k_i$ denotes the degree of vertex $i$ so $k_i=\sum_{j}A_{i,j}$.  Further denote by $\mathbf s\in \{1,\ldots,n\}^n$  a vector having the same dimension of $A$, and representing an allocation of vertexes to communities. Thus, $\mathbf s_i$ represents the community assigned to vertex $i$.  Now the idea is to compare the graph $G$ with a graph $G'$ having no community structure.  A group $V_k=\{i \in V \mid \mathbf s_{i}=k\}$ possess an accumulated weight of $\sum_{i,j \in  V_k} A_{i,j}$. Now for $G'$, assuming it is a random instance of an Erdős-R\'enyi graph,   the set $V_k$ should have an accumulated weight of $\sum_{i,j \in V_k} \frac{k_ik_j}{2m}$. Hence, the difference $\sum_{i,j \in  V_k} A_{i,j}-\frac{k_ik_j}{2m}$ quantifies how distant is the immersion of community $V_k$ in the graph $G$ from $G'$. The modularity function is now defined as the sum over all communities:
\[
Q(\mathbf s) := \sum_{k} \sum_{i,j \in  V_k} \left( A_{i,j}-\frac{k_ik_j}{2m}\right) 
=
\sum_{i,j \in  V} \left( A_{i,j}-\frac{k_ik_j}{2m}\right)  \delta(\mathbf s_i,\mathbf s_j),
\]
where  $\delta(\mathbf s_i,\mathbf s_j)=0$ unless $\mathbf s_i=\mathbf s_j$ in which case $\delta(\mathbf s_i,\mathbf s_j)=1$.

As such, the modularity function $Q(\cdot)$ is defined for unweighted undirected graphs. In particular, for graphs obtained from a correlation matrix, which indeed is weighted, the modularity function $Q(\cdot)$ requires to be adjusted. Moreover, the null model (the graph $G'$) is critical for the well-functioning of modularity; see e.g., the discussion in \cite{Fortunato2016}. Hence, to couple with this problem, we choose to work with the formulation of \cite{MacMahon2015} where correlation matrix is filtered and modularity is adjusted for the right ``null model'' $G'$.  The analysis is again based on a spectral analysis as we now explain. Let $C$ be a correlation matrix and consider the set of eigenvalues $\lambda_1,\ldots, \lambda_n$ which we assume are displayed in increasing order. Let $v_1,\ldots,v_n$ be the corresponding eigenvectors. Moreover, let $T$ be the number of observations and the critical values
\[
\lambda_{-}:=\left( 1 - \sqrt{\frac{n}{T}}\right)^2, \hspace{1cm}  \lambda_{+}:=\left( 1+ \sqrt{\frac{n}{T}}\right)^2.
\]
The values $\lambda_-,\lambda_+$ are parameters for  Marcenko-Pastur distribution in random matrix theory which is given by $\rho(\lambda)=\frac{T}{n} \frac{\sqrt{(\lambda_+-\lambda)(\lambda-\lambda_+)}}{2 \pi \lambda}$. Define the matrices
\begin{align}
C^{r}&:=\sum_{\lambda_i \leq  \lambda_+} \lambda_{i} v_i^{\tr} \cdot  v_i\\
C^{g}&:=\sum_{\lambda_+ < \lambda_i < \lambda_n} \lambda_{i} v_i^{\tr} \cdot  v_i\\
C^{m}&:=\lambda_{n} v_n^{\tr} \cdot  v_n.
\end{align}
We have a decomposition of the correlation matrix $C$ given by
\begin{equation}\label{eq:decompositionofC}
C=C^{m}+ C^{g}  + C^{r}.
\end{equation}
From the ordering of the eigenvalues, the matrix $C^{r}$ represents some random noise, $C^{m}$ a global signal which in our financial context is attached to the market as a whole and $C^{g}$ represents information in a mesoscopic scale just between $C^r$ and $C^m$. However, the set of eigenvalues $\lambda_i$ satisfying $\lambda_+ < \lambda_i < \lambda_n$ could be empty (as we will find) and in this case there makes no sense to consider $C^g$.  Next, we explain how the modularity function $Q(\cdot)$ is adjusted. Accordingly,  focusing in the matrix $C^g$, and taking into account the decomposition \eqref{eq:decompositionofC}, the null model is $C^r + C^m$ and the modularity functions takes the form
\begin{align}\label{eq:modularity3}
Q_3(\mathbf s):=&\frac{1}{C_{norm}}\sum_{i,j} [C_{i,j} - C^r_{i,j} -C^m_{i,j}]  \delta(\mathbf s_i,\mathbf s_j)\notag\\
=&\frac{1}{C_{norm}}\sum_{i,j} C^g_{i,j}  \delta(\mathbf s_i,\mathbf s_j)
\end{align}
for $C_{norm}=\sum_{i,j}C_{i,j}$ a normalizing constant. However, as we mentioned before, for some empirical correlation matrices, the matrix $C^g$ will be null. Hence, it also makes sense to consider a decomposition $C=C^s + C^r$ with $C^s:=\sum_{\lambda_+ < \lambda_i} \lambda_{i} v_i^{\tr} \cdot  v_i\\$ and then the modularity is defined by
\begin{align}\label{eq:modularity2}
Q_2(\mathbf s):=&\frac{1}{C_{norm}}\sum_{i,j} [C_{i,j} - C^r_{i,j}]  \delta(\mathbf s_i,\mathbf s_j)\notag\\
=&\frac{1}{C_{norm}}\sum_{i,j} C^s_{i,j}  \delta(\mathbf s_i,\mathbf s_j).
\end{align}

Hence, in this section we maximize the modularity functions $Q_2$ and $Q_3$ in order to define communities and report on them. It is known that the maximization of modularity functions is a  \textit{NP-hard} problem; see \cite{Brandes2006}. Hence, the optimization is approached through several heuristic algorithms. We implement the popular Louvian algorithm, adjusted as described by \cite{MacMahon2015} accordingly to the modularity functions $Q_2$ and $Q_3$. 

\subsection{Modularity function $Q_2$}\label{sec:modularityQ2}
\begin{figure}[H]
\centering
\includegraphics[width=12cm]{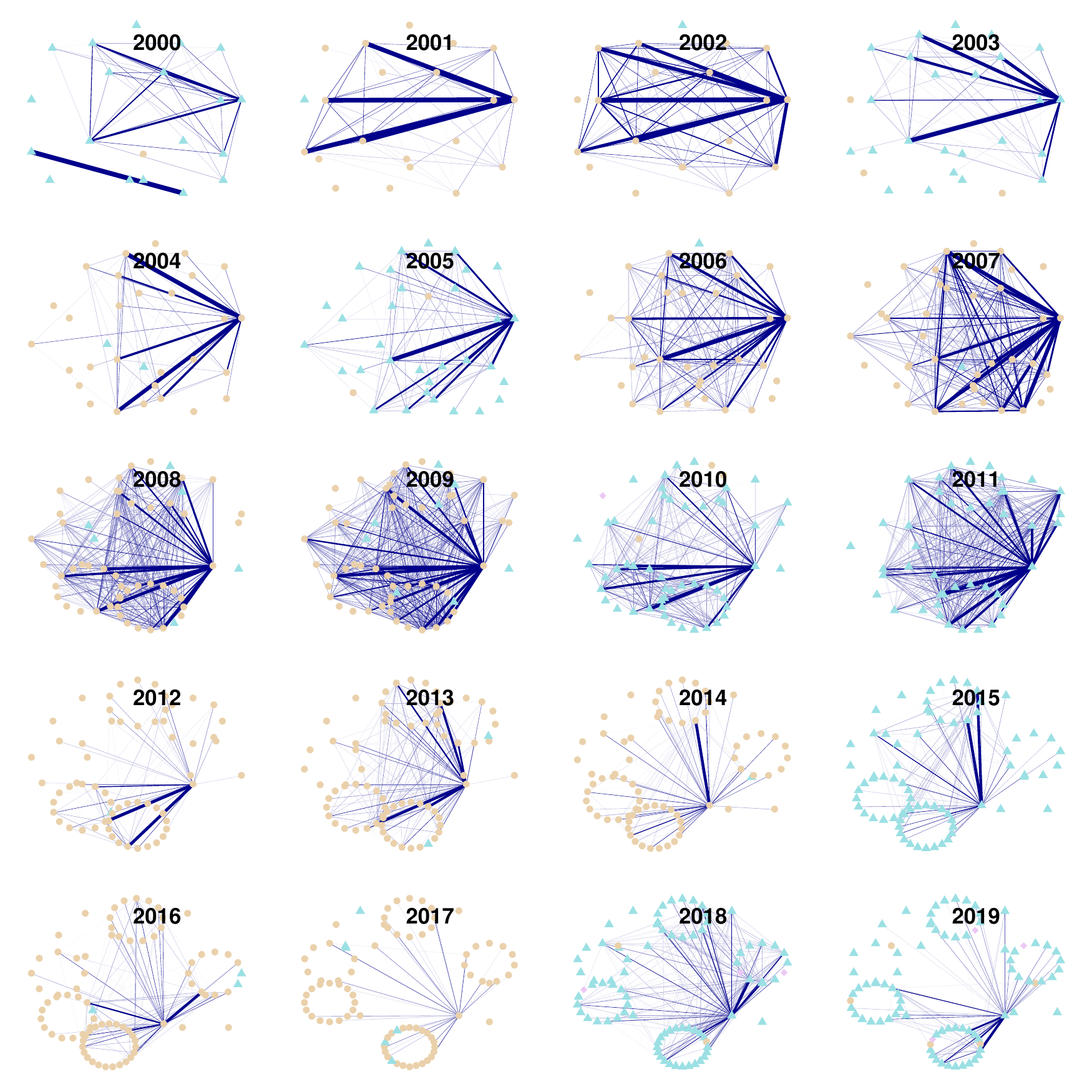}
\caption{Communities obtained from modularity function $Q_2$. The color of the nodes represent community, which is equivalently represented by vertex' shape. For visuality only edges with weights in absolute value in the interval $[.3,\infty)$ are shown. Only weights above $0.5$ in absolute value are distinguished in the represented edge's width.}
\label{figcommunitesQ2}
\end{figure}

In Figure \ref{figcommunitesQ2} we see the resulting communities obtained with the Louvian algorithm applied to the modularity function $Q_2$ defined in  \eqref{eq:modularity2}. In all of the years of the period there are two communities. The first community  is a ``giant component'' and the other community consist of a small number of isolated vertexes.  Hence, at this scale our procedure does not detect a complex community structure. This is what we expected. Note however the stylized fact:
\begin{itemize}
\item The turmoil at the Subprime crisis period is captured by a visually evident increase on interconnectedness particularly for the years 2007 to 2009. 
\end{itemize}

\subsection{Modularity function $Q_3$}\label{sec:modularityQ3}
\begin{figure}[H]
\centering
\includegraphics[width=6cm]{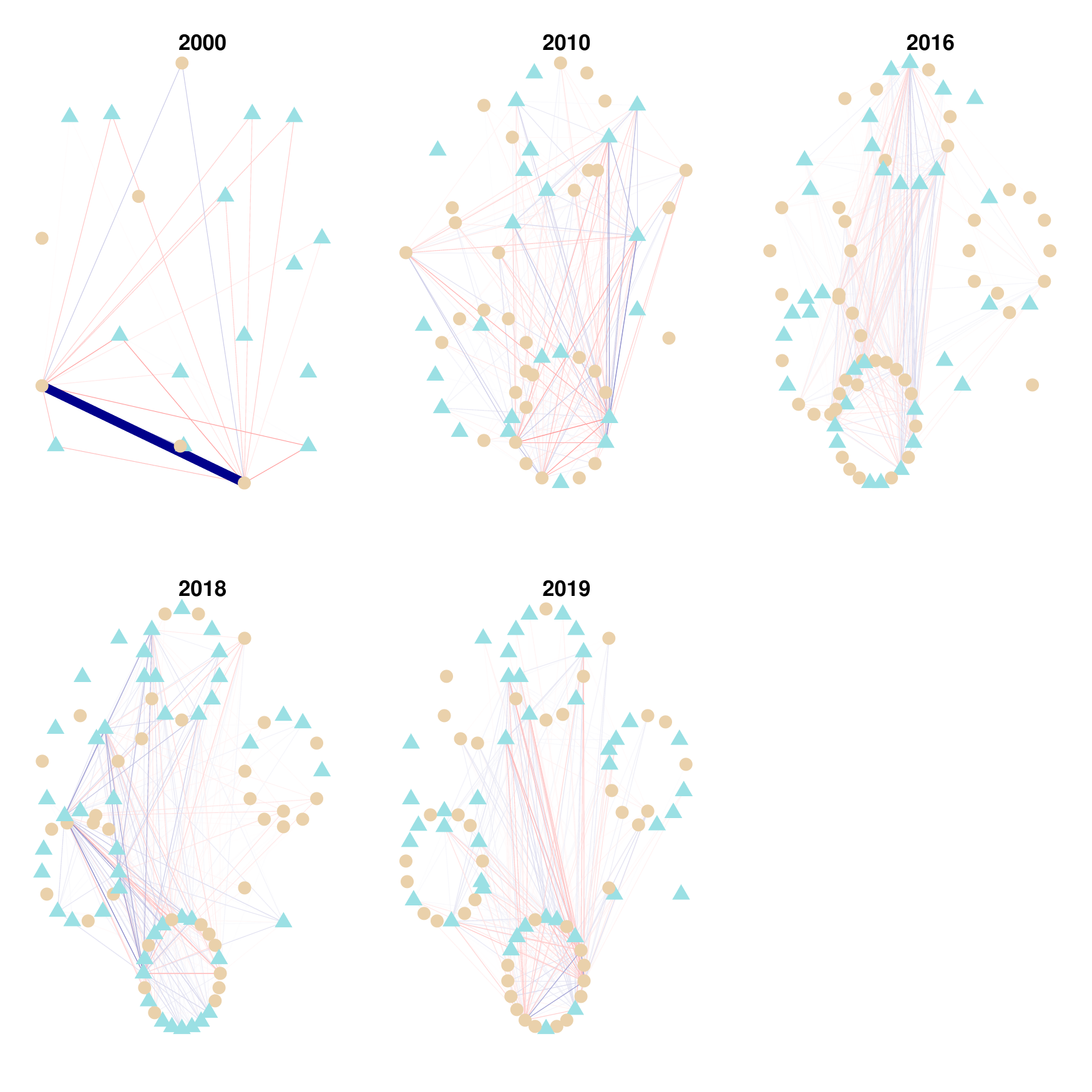}
\caption{Communities from modularity function $Q_3$. The color of nodes identifies membership to the same community and equivalently for the vertex' shape. For visibility only edges with weights in absolute value in the intervals $[.05,\infty)$ are shown. Only weights above $0.5$ in absolute value are distinguished in the represented edge's width.}
\label{Fig:70communitiesQ3}
\end{figure}

For the definition of the modularity function $Q_3$ the matrix $C^g$ is necessary and should not be a null matrix.  For our data, this is the case for only a few years: 2000, 2010, 2016, 2018 and 2019. For these years, a representation of communities can be seen from Figure \ref{Fig:70communitiesQ3}.  This is what we observe:  
\begin{itemize}
\item First of all, in each year, there are only two communities as can be seen from the color of the vertexes, or equivalently from their shape. Interestingly, there is no clear larger community.
\item Second, and also interesting, for our data, industrial sector is non determinant for the community assignment. More clearly, each industrial sector have vertexes in each community. This fact should be compared with the finding of Section \ref{sec:GGmStylf} based on partial correlations where there also existed intersectorial links.
\item This is our explanation of the years in which there existed a non trivial matrix $C^g$. First of all recall that this matrix represents structure between individual stocks and the market as a whole, while in crisis periods this last structure is what prevails since stocks tend to be highly correlated at those times. In the year 2000 we find the peak (and burst) of the dot com bubble  from which for the years 2001 and 2002  bearish markets prevailed.  What we see from Figure \ref{figcommunitesQ2} for the network constructed from the matrices $C^s$ is an increase in interconnectedness while in Figure \ref{Fig:70communitiesQ3} we see that in the period 2000:2002 there existed a ``mesoscopic'' structure for the year 2000 in which there is a ``local minimum'' for graphs interconnectedness. Analogously for the year 2010 in Figure \ref{Fig:70communitiesQ3} which coincides with a local minimum in Figure \ref{figcommunitesQ2} for the ``extended'' subprime crisis period 2007-2010.
\item Now we compare the years 2016, 2018 and 2019 in Figures \ref{figcommunitesQ2} and \ref{Fig:70communitiesQ3}.  Those are years in which global events occurred, to mention some of them: The Brexit (starting from its referendum in june 2016), US elections for the period 2017-2020, China-US trade war starting from july 2018.  However, none of these seems to be comparable to the levels of the dot.com bubble and the subprime crisis.  In particular for the Mexican stock market they didn't have such an impact as to hide the effects of a mesoscopic structure and inducing all stocks as moving according to a unique factor.
\end{itemize}

\section{Conclusion}\label{sec:FinancialDiscussion}
In global crisis periods, price levels of stocks in the Mexican stock exchange indeed present obvious changes  which are visually evident and confirmed by econometric models as we have shown here and is also documented by other authors.  However, the interdependency structure is a more complex phenomenon and much less studied.  Our findings show that as long as partial correlations are concerned, the interdependency structure is quite stable and only centrality metrics from network theory have the fine sensibility to quantify changes. Degree and eigen centralities indeed present a discontinuous variation, an upwards jump at the peak of the crisis and then a downwards jump when the shock of the crisis has been absorbed in the market. Another interesting finding of studying interdependency structure from partial correlations is that only a small number of negative partial correlations which are also in magnitude small are present. We argue this is an indicator of a positive synergy of an integrated market.  Reinforcing this claim, we find that industrial sectors are strongly interconnected even at the level of partial correlations, which is a less established property. In general and in particular for the Mexican case.\\

Interdependency from the point of view of (``absolute'') correlations confirms findings from partial correlations. It also provides evidence of an integrated market for the Mexican case.  Indeed, this is what we learned from the estimation of modularities which determined community structure with no separation of industrial sector.  From filtered matrices with noise filtered out (the matrices $C^s$) a single giant component emerged. Moreover, here the effect of global episodes for interdependency structure was quite clear even for visual appreciation. This is what we learned in Figure \ref{figcommunitesQ2} and is perfect as evidence for the  modeling strength.  Indeed, correlations are more sensible to trading activity than partial correlations, and capture relationships among stocks due to such activity which is even more pronounced at crisis periods. We also studied community structure from the matrices $C^g$ which are the correlation matrices after noise and the global market mode have been filtered out. At this scale it happens that only a few observed years present a mesoscopic structure.  For the years 2000 and 2010 in which mesoscopic structure is  present, we observe a ``local minimum'' for interconnecctedness in Figure \ref{figcommunitesQ2}. For the years 2016, 2018, 2019  we note also a turmoil of stress periods (e.g., the Brexit, China-US trade affair, etc) which nevertheless are not to be compared in serverity with the episodes of the dot.com bubble and the subprime crisis so are not able to blur the presence of structure at the mesoscopic level.

\appendix
\section{Filtered list of stocks}\label{sec:Selectedstocks}
In this appendix we ilustrate the collection of stocks selected for each year in the period 2000-2019.

\begin{figure}[H]
\begin{center}
	\includegraphics[width=15cm]{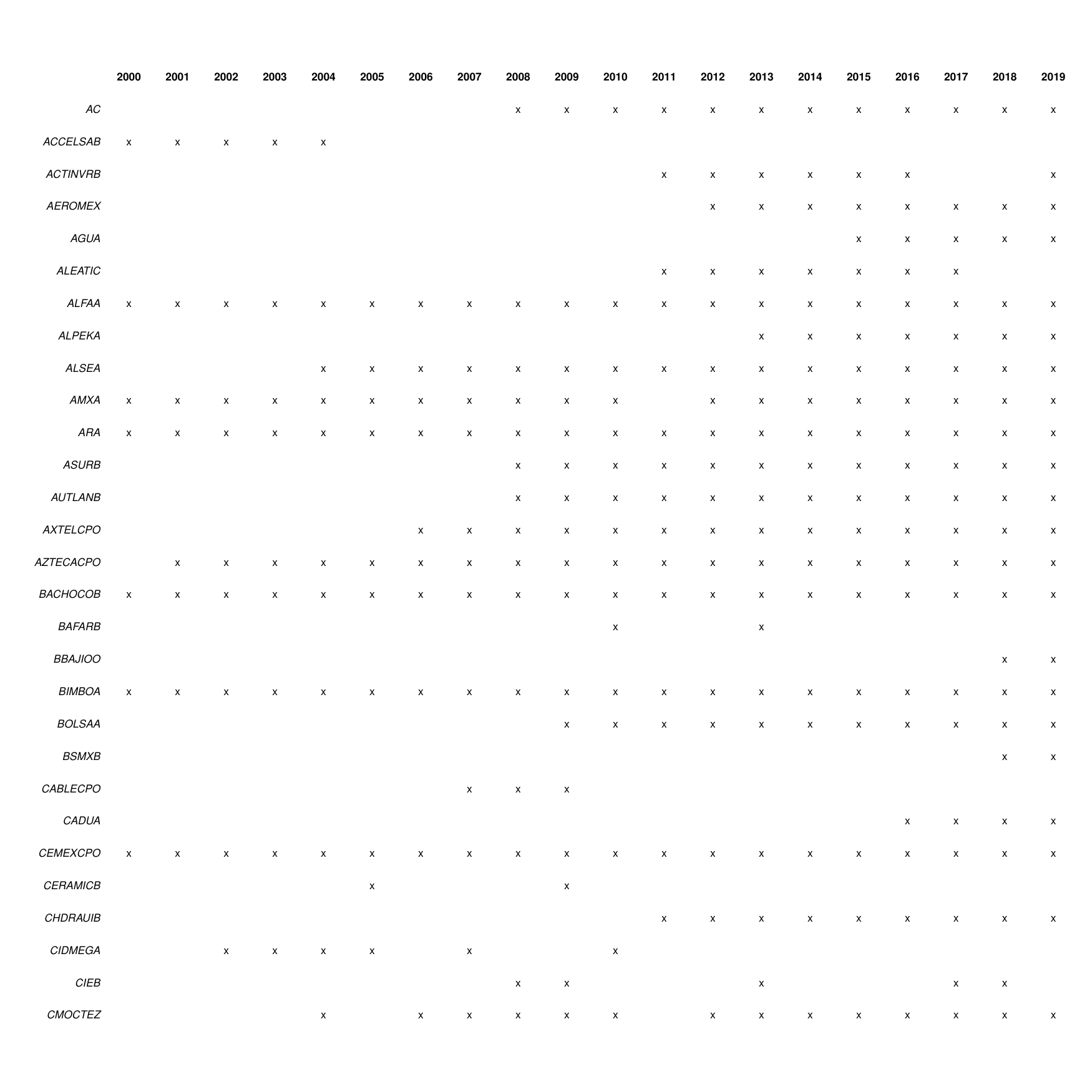}\label{table.1}
		\caption{Filtered list of stocks 1/4.}
\end{center}
\end{figure}

\begin{figure}[H]
\begin{center}
	\includegraphics[width=15cm]{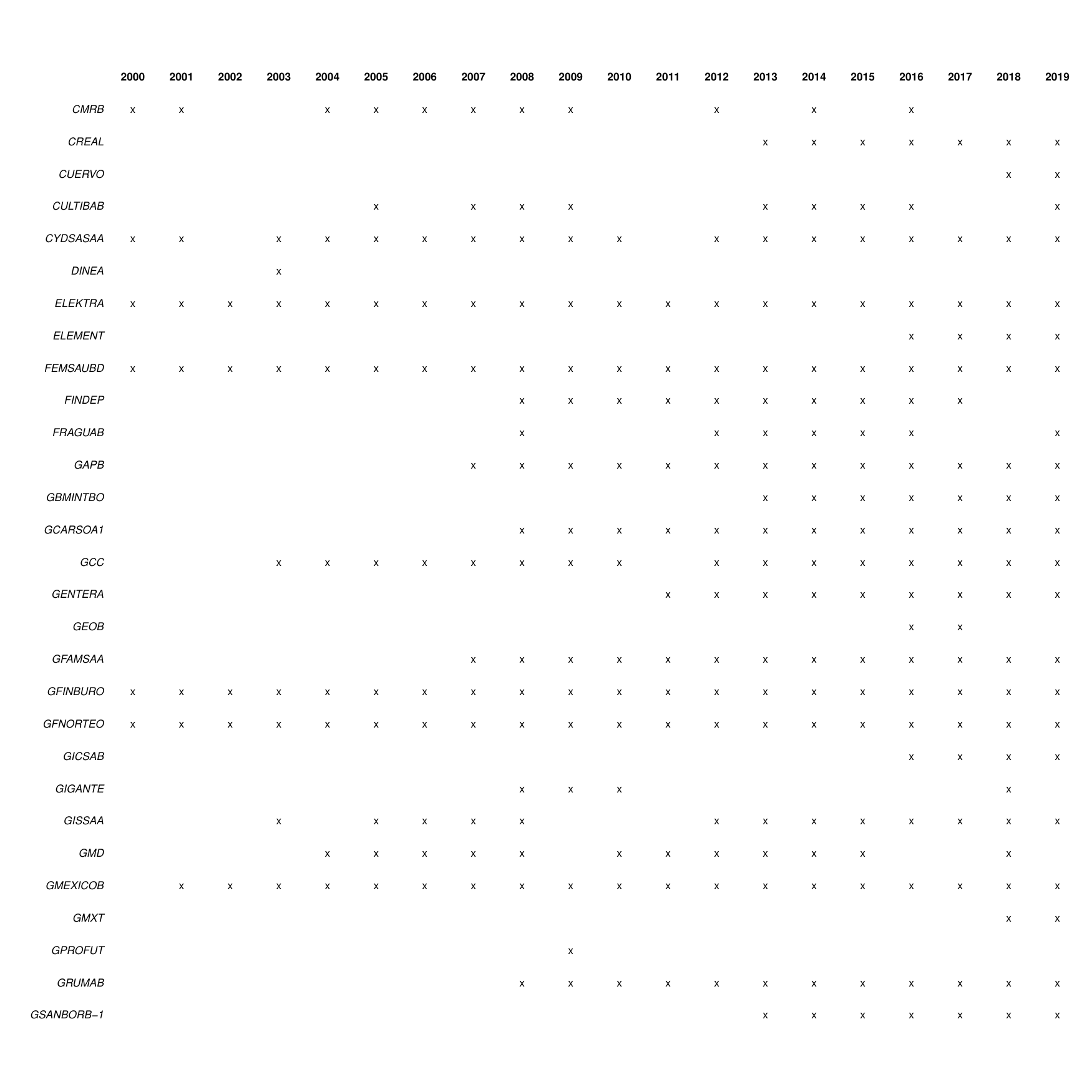}\label{table.2}
		\caption{Filtered list of stocks 2/4.}
\end{center}
\end{figure}

\begin{figure}[H]
\begin{center}
	\includegraphics[width=15cm]{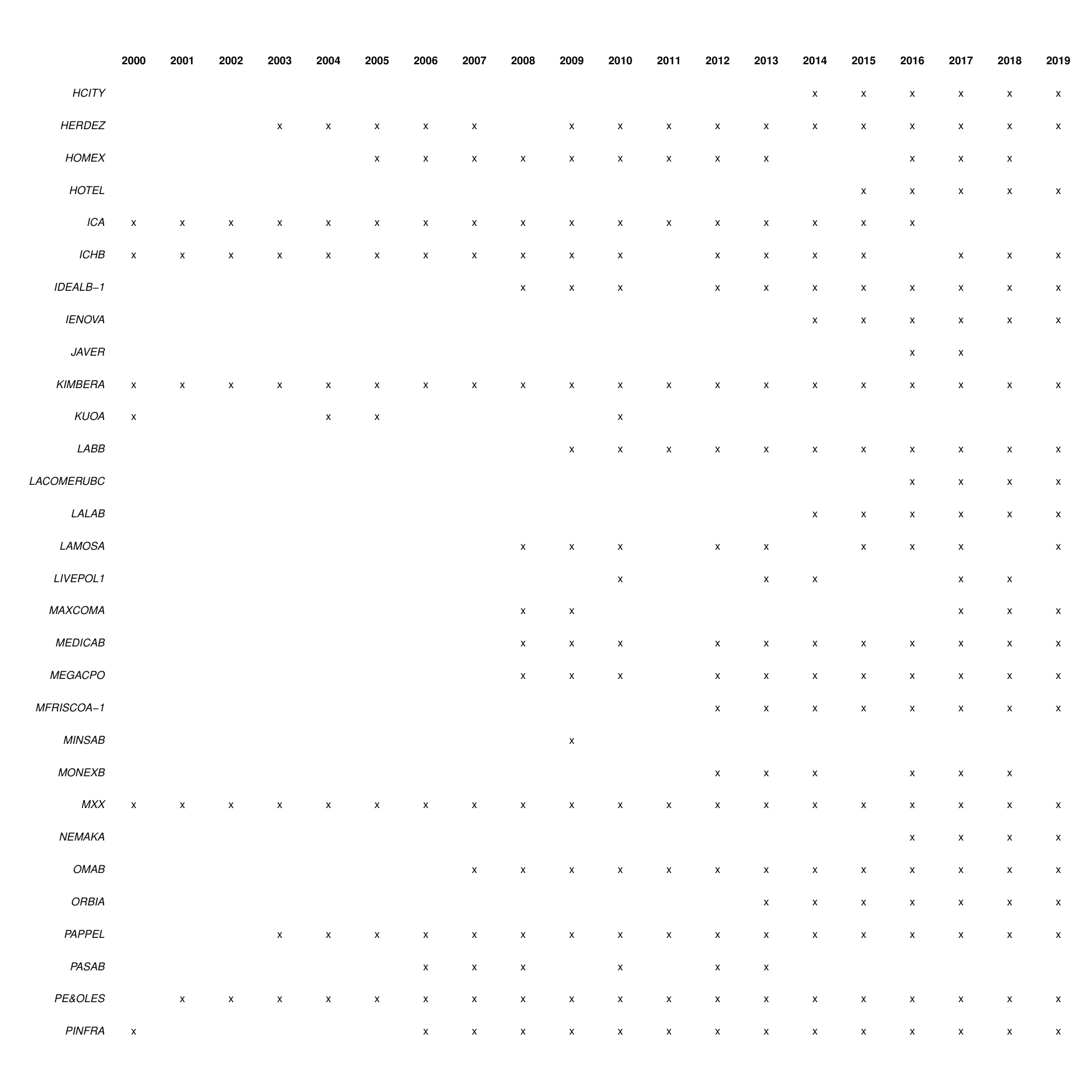}\label{table.3}
		\caption{Filtered list of stocks 3/4.}
\end{center}
\end{figure}

\begin{figure}[H]
	\begin{center}
\includegraphics[width=15cm]{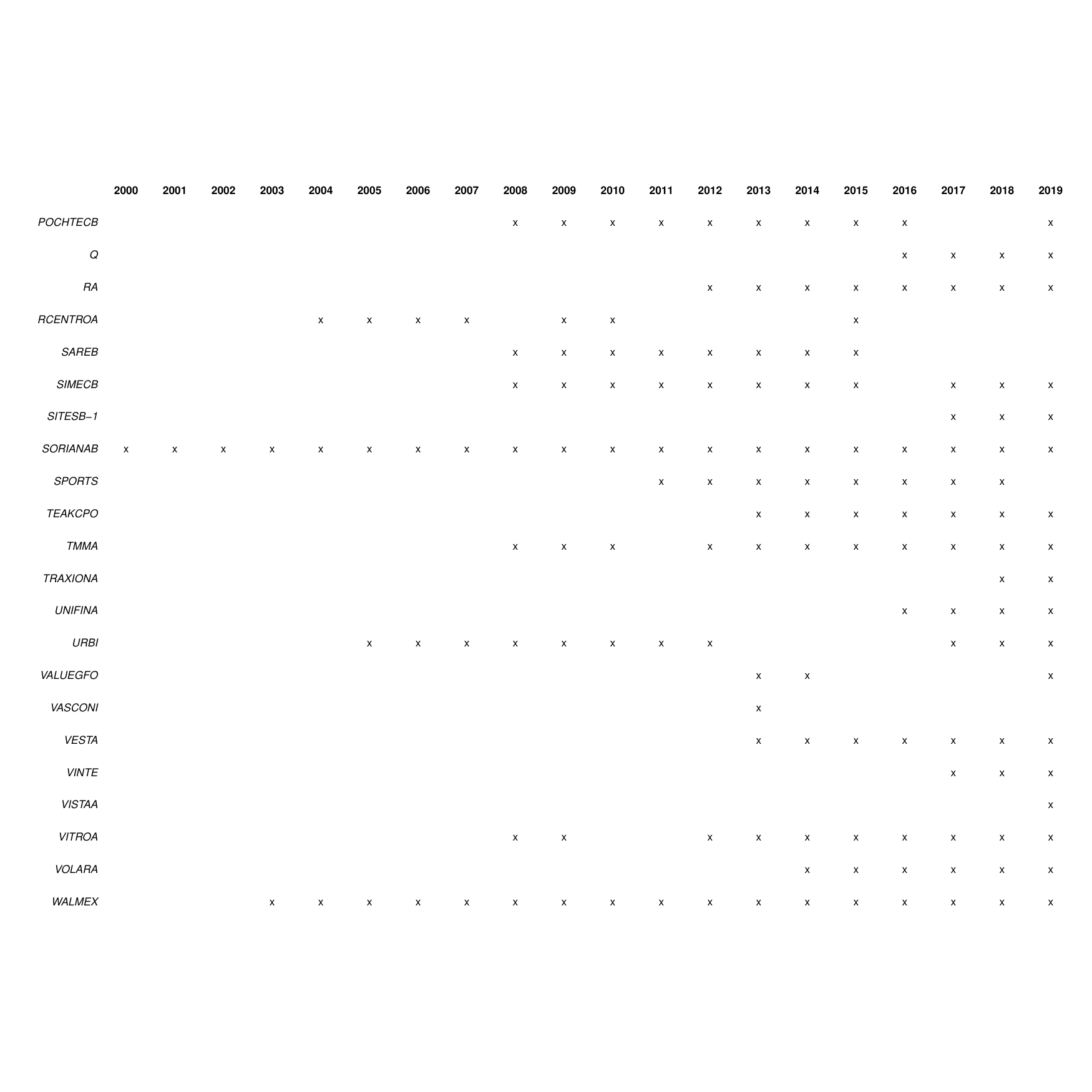}\label{table.4}
\caption{Filtered list of stocks 4/4.}
\end{center}
\end{figure}

\section{Partial correlations}\label{sec:Partialcorrelations}
In Table \ref{tab:linksp3yuno} we display partial correlations above the threshold 0.3 in absolute value. 
\begin{table}[H]
\caption{Links in the rank $[0.3,1]$ for the period 2000-2009.}
\label{tab:linksp3yuno}
\centering
\begin{adjustbox}{width=\columnwidth,center}
\begin{tabular}{llllllll}
	\hline
	tick1 & tick2 & Par.Corr & year & tick1 & tick2 & Par.Corr & year \\ 
	\hline
	ELEKTRA & ICA & 0.98 & 2000 & HOMEX & URBI & 0.32 & 2012 \\ 
	FEMSAUBD & MXX & 0.4 & 2000 & ICHB & SIMECB & 0.43 & 2012 \\ 
	AZTECACPO & SORIANAB & 0.33 & 2001 & AMXA & MXX & 0.87 & 2013 \\ 
	AMXA & MXX & 0.37 & 2002 & FEMSAUBD & MXX & 0.44 & 2013 \\ 
	AZTECACPO & MXX & 0.38 & 2002 & GMEXICOB & MXX & 0.48 & 2013 \\ 
	CEMEXCPO & MXX & 0.38 & 2002 & ICHB & SIMECB & 0.34 & 2013 \\ 
	FEMSAUBD & MXX & 0.42 & 2002 & MXX & WALMEX & 0.31 & 2013 \\ 
	MXX & SORIANAB & 0.32 & 2002 & CEMEXCPO & MXX & 0.33 & 2014 \\ 
	AMXA & MXX & 0.57 & 2003 & FEMSAUBD & MXX & 0.59 & 2014 \\ 
	AZTECACPO & MXX & 0.32 & 2003 & ASURB & GAPB & 0.35 & 2015 \\ 
	CEMEXCPO & MXX & 0.36 & 2003 & CEMEXCPO & MXX & 0.33 & 2015 \\ 
	FEMSAUBD & MXX & 0.34 & 2003 & FEMSAUBD & MXX & 0.42 & 2015 \\ 
	MXX & WALMEX & 0.76 & 2003 & GFNORTEO & MXX & 0.37 & 2015 \\ 
	AMXA & MXX & 0.7 & 2004 & GMEXICOB & MXX & 0.39 & 2015 \\ 
	CEMEXCPO & MXX & 0.52 & 2004 & ICHB & SIMECB & 0.36 & 2015 \\ 
	MXX & WALMEX & 0.43 & 2004 & CEMEXCPO & MXX & 0.46 & 2016 \\ 
	CEMEXCPO & MXX & 0.44 & 2005 & FEMSAUBD & MXX & 0.61 & 2016 \\ 
	MXX & WALMEX & 0.57 & 2005 & GFNORTEO & MXX & 0.36 & 2016 \\ 
	CEMEXCPO & MXX & 0.5 & 2006 & MXX & WALMEX & 0.34 & 2016 \\ 
	FEMSAUBD & MXX & 0.42 & 2006 & ASURB & GAPB & 0.33 & 2017 \\ 
	GMEXICOB & MXX & 0.37 & 2006 & CEMEXCPO & MXX & 0.78 & 2017 \\ 
	MXX & WALMEX & 0.83 & 2006 & FEMSAUBD & MXX & 0.57 & 2017 \\ 
	CEMEXCPO & MXX & 0.5 & 2007 & GEOB & HOMEX & 0.36 & 2017 \\ 
	GAPB & OMAB & 0.38 & 2007 & GFNORTEO & MXX & 0.46 & 2017 \\ 
	GMEXICOB & MXX & 0.31 & 2007 & GMEXICOB & MXX & 0.3 & 2017 \\ 
	MXX & WALMEX & 0.58 & 2007 & ICHB & SIMECB & 0.37 & 2017 \\ 
	ALFAA & ARA & 0.3 & 2008 & ASURB & GAPB & 0.33 & 2018 \\ 
	GAPB & OMAB & 0.31 & 2008 & CEMEXCPO & MXX & 0.74 & 2018 \\ 
	ICHB & SIMECB & 0.41 & 2008 & FEMSAUBD & MXX & 0.84 & 2018 \\ 
	MXX & WALMEX & 0.52 & 2008 & GFNORTEO & MXX & 0.65 & 2018 \\ 
	CEMEXCPO & MXX & 0.37 & 2009 & GIGANTE & LIVEPOL1 & 0.3 & 2018 \\ 
	GAPB & OMAB & 0.34 & 2009 & MXX & WALMEX & 0.37 & 2018 \\ 
	ICHB & SIMECB & 0.34 & 2009 & ASURB & GAPB & 0.31 & 2019 \\ 
	MXX & WALMEX & 0.49 & 2009 & CEMEXCPO & MXX & 0.58 & 2019 \\ 
	CEMEXCPO & MXX & 0.35 & 2010 & FEMSAUBD & GFNORTEO & -0.33 & 2019 \\ 
	ICHB & SIMECB & 0.34 & 2010 & FEMSAUBD & MXX & 0.79 & 2019 \\ 
	MXX & WALMEX & 0.36 & 2010 & GAPB & OMAB & 0.33 & 2019 \\ 
	FEMSAUBD & MXX & 0.35 & 2011 & GFNORTEO & MXX & 0.83 & 2019 \\ 
	MXX & WALMEX & 0.33 & 2011 & GMEXICOB & MXX & 0.54 & 2019 \\ 
	\hline
\end{tabular}
\end{adjustbox}
\end{table}


Partial correlations in absolute value in the interval $[0.2,0.3]$ are displayed in Table \ref{tab:linksp2yp3}. 

\begin{table}[H]
\caption{Links in the rank $[0.2,0.3]$ for the period 2000-2009.}
\label{tab:linksp2yp3}
\centering
\begin{adjustbox}{width=\columnwidth,center}
\begin{tabular}{llllllll}
	\hline
	tick1 & tick2 & Par.Corr & year & tick1 & tick2 & Par.Corr & year \\ 
	\hline
	CEMEXCPO & MXX & 0.26 & 2000 & GCARSOA1 & GFINBURO & 0.26 & 2010 \\ 
	GFINBURO & MXX & 0.21 & 2000 & GFNORTEO & MXX & 0.21 & 2010 \\ 
	GFNORTEO & MXX & 0.24 & 2000 & GMEXICOB & MXX & 0.29 & 2010 \\ 
	MXX & SORIANAB & 0.25 & 2000 & HOMEX & ICA & 0.21 & 2010 \\ 
	ARA & GFNORTEO & 0.25 & 2001 & ALFAA & MXX & 0.22 & 2011 \\ 
	AZTECACPO & ELEKTRA & 0.21 & 2001 & CEMEXCPO & ICA & 0.27 & 2011 \\ 
	CEMEXCPO & FEMSAUBD & 0.29 & 2001 & CEMEXCPO & MXX & 0.21 & 2011 \\ 
	ALFAA & MXX & 0.2 & 2002 & ELEKTRA & MXX & 0.22 & 2011 \\ 
	AZTECACPO & ELEKTRA & 0.22 & 2002 & GFNORTEO & MXX & 0.3 & 2011 \\ 
	GFINBURO & MXX & 0.29 & 2002 & GMEXICOB & MXX & 0.29 & 2011 \\ 
	GFNORTEO & MXX & 0.23 & 2002 & HOMEX & URBI & 0.28 & 2011 \\ 
	ARA & MXX & 0.22 & 2003 & CEMEXCPO & MXX & 0.27 & 2012 \\ 
	GFINBURO & MXX & 0.21 & 2003 & FEMSAUBD & MXX & 0.28 & 2012 \\ 
	MXX & SORIANAB & 0.26 & 2003 & GMEXICOB & MXX & 0.26 & 2012 \\ 
	ALFAA & MXX & 0.29 & 2004 & MXX & WALMEX & 0.22 & 2012 \\ 
	AMXA & CEMEXCPO & -0.23 & 2004 & ALFAA & MXX & 0.27 & 2013 \\ 
	AZTECACPO & MXX & 0.2 & 2004 & AMXA & FEMSAUBD & -0.24 & 2013 \\ 
	GFINBURO & MXX & 0.24 & 2004 & CEMEXCPO & MXX & 0.22 & 2013 \\ 
	GMEXICOB & MXX & 0.26 & 2004 & CULTIBAB & TEAKCPO & -0.22 & 2013 \\ 
	GMEXICOB & PE\&OLES & 0.22 & 2004 & GFNORTEO & MXX & 0.26 & 2013 \\ 
	MXX & SORIANAB & 0.23 & 2004 & GMEXICOB & PE\&OLES & 0.23 & 2013 \\ 
	ALFAA & MXX & 0.23 & 2005 & HOMEX & SAREB & 0.22 & 2013 \\ 
	AMXA & MXX & 0.21 & 2005 & ALFAA & MXX & 0.24 & 2014 \\ 
	ARA & URBI & 0.21 & 2005 & ALSEA & CULTIBAB & 0.25 & 2014 \\ 
	FEMSAUBD & MXX & 0.22 & 2005 & ASURB & GAPB & 0.21 & 2014 \\ 
	GMEXICOB & MXX & 0.25 & 2005 & GFNORTEO & MXX & 0.26 & 2014 \\ 
	KIMBERA & MXX & 0.2 & 2005 & GMEXICOB & MXX & 0.3 & 2014 \\ 
	ALFAA & MXX & 0.23 & 2006 & MFRISCOA-1 & PE\&OLES & 0.27 & 2014 \\ 
	AMXA & MXX & 0.25 & 2006 & ALFAA & MXX & 0.21 & 2015 \\ 
	GFINBURO & MXX & 0.27 & 2006 & GFINBURO & MXX & 0.22 & 2015 \\ 
	GFNORTEO & MXX & 0.23 & 2006 & MXX & WALMEX & 0.2 & 2015 \\ 
	GMEXICOB & WALMEX & -0.2 & 2006 & AC & BIMBOA & 0.21 & 2016 \\ 
	MXX & PINFRA & 0.21 & 2006 & ALFAA & GFINBURO & 0.22 & 2016 \\ 
	BIMBOA & MXX & 0.23 & 2007 & ASURB & GAPB & 0.2 & 2016 \\ 
	GFNORTEO & MXX & 0.23 & 2007 & GENTERA & PINFRA & 0.23 & 2016 \\ 
	HOMEX & MXX & 0.2 & 2007 & GFNORTEO & RA & 0.21 & 2016 \\ 
	ICA & MXX & 0.24 & 2007 & MFRISCOA-1 & PE\&OLES & 0.26 & 2016 \\ 
	MXX & URBI & 0.23 & 2007 & MXX & ORBIA & 0.2 & 2016 \\ 
	ALFAA & AXTELCPO & 0.25 & 2008 & ALFAA & ALPEKA & 0.24 & 2017 \\ 
	AMXA & MXX & 0.23 & 2008 & CEMEXCPO & FEMSAUBD & -0.27 & 2017 \\ 
	CEMEXCPO & MXX & 0.24 & 2008 & CEMEXCPO & GFNORTEO & -0.24 & 2017 \\ 
	FEMSAUBD & MXX & 0.26 & 2008 & GAPB & OMAB & 0.2 & 2017 \\ 
	GCARSOA1 & MXX & 0.21 & 2008 & GFNORTEO & GMEXICOB & -0.21 & 2017 \\ 
	GMEXICOB & MXX & 0.28 & 2008 & HOMEX & URBI & 0.27 & 2017 \\ 
	MXX & PE\&OLES & 0.25 & 2008 & MXX & WALMEX & 0.24 & 2017 \\ 
	BIMBOA & MXX & 0.21 & 2009 & CEMEXCPO & FEMSAUBD & -0.26 & 2018 \\ 
	FEMSAUBD & MXX & 0.21 & 2009 & FEMSAUBD & GFNORTEO & -0.25 & 2018 \\ 
	GCARSOA1 & MXX & 0.21 & 2009 & GAPB & OMAB & 0.23 & 2018 \\ 
	GMEXICOB & MXX & 0.26 & 2009 & GMEXICOB & MXX & 0.26 & 2018 \\ 
	HOMEX & MXX & 0.25 & 2009 & ICHB & SIMECB & 0.25 & 2018 \\ 
	ASURB & GAPB & 0.24 & 2010 & CEMEXCPO & FEMSAUBD & -0.24 & 2019 \\ 
	AXTELCPO & GFAMSAA & 0.21 & 2010 & CEMEXCPO & WALMEX & -0.21 & 2019 \\ 
	\hline
\end{tabular}
\end{adjustbox}
\end{table}

%

\section{Persistent links from partial correlations}\label{appendix:Endurablelinks}

\begin{table}[H]
	\caption{Persistent links in the rank $[0.1,1]$ for the period 2000-2009 (1/2).}
	\centering
	\begin{adjustbox}{width=\columnwidth,center}
		\begin{tabular}{rrrrrrrrrrr}
			\hline
			& 2000 & 2001 & 2002 & 2003 & 2004 & 2005 & 2006 & 2007 & 2008 & 2009 \\ 
			\hline
			ALFAA-MXX & 0.17 &  & 0.21 &  & 0.29 & 0.24 & 0.24 &  &  &  \\ 
			AMXA-MXX & 0.16 &  & 0.37 & 0.61 & 0.69 & 0.21 & 0.28 &  & 0.26 & 0.21 \\ 
			CEMEXCPO-MXX & 0.27 & 0.14 & 0.39 & 0.36 & 0.51 & 0.45 & 0.48 & 0.49 & 0.26 & 0.37 \\ 
			FEMSAUBD-MXX & 0.40 & 0.17 & 0.45 & 0.34 & 0.19 & 0.23 & 0.44 & 0.19 & 0.26 & 0.21 \\ 
			GFINBURO-MXX & 0.21 & 0.15 & 0.28 & 0.21 & 0.24 &  & 0.26 &  &  & 0.17 \\ 
			GFNORTEO-MXX & 0.24 &  & 0.24 &  &  &  & 0.20 & 0.24 &  & 0.20 \\ 
			BIMBOA-MXX &  &  & 0.17 &  & 0.19 & 0.17 &  & 0.23 & 0.15 & 0.22 \\ 
			MXX-WALMEX &  &  &  & 0.77 & 0.43 & 0.57 & 0.77 & 0.54 & 0.37 & 0.56 \\ 
			GMEXICOB-MXX &  &  &  &  & 0.26 & 0.26 & 0.34 & 0.32 & 0.27 & 0.26 \\ 
			ASURB-GAPB &  &  &  &  &  &  &  &  & 0.17 & 0.13 \\ 
			ICHB-SIMECB &  &  &  &  &  &  &  &  & 0.41 & 0.34 \\ 
			\hline
		\end{tabular}
	\end{adjustbox}
\end{table}

\begin{table}[H]
	\caption{Persistent links in the rank $[0.1,1]$ for the period 2010-2019 (2/2).}
	\centering
	\begin{adjustbox}{width=\columnwidth,center}
		\begin{tabular}{rrrrrrrrrrr}
			\hline
			& 2010 & 2011 & 2012 & 2013 & 2014 & 2015 & 2016 & 2017 & 2018 & 2019 \\ 
			\hline
			ALFAA-MXX &  & 0.23 & 0.20 & 0.29 & 0.24 & 0.22 &  &  &  &  \\ 
			AMXA-MXX &  &  & 0.11 & 0.85 &  &  &  &  &  &  \\ 
			CEMEXCPO-MXX & 0.36 & 0.21 & 0.27 & 0.22 & 0.33 & 0.36 & 0.45 & 0.72 & 0.68 & 0.55 \\ 
			FEMSAUBD-MXX &  & 0.35 & 0.25 & 0.47 & 0.48 & 0.41 & 0.62 & 0.50 & 0.81 & 0.72 \\ 
			GFINBURO-MXX & 0.14 &  &  &  &  & 0.22 &  &  &  & 0.13 \\ 
			GFNORTEO-MXX & 0.21 & 0.30 & 0.19 & 0.26 & 0.26 & 0.37 & 0.37 & 0.47 & 0.63 & 0.81 \\ 
			BIMBOA-MXX & 0.20 &  &  &  & 0.16 & 0.20 & 0.14 & 0.13 & 0.17 &  \\ 
			MXX-WALMEX & 0.39 & 0.33 & 0.22 & 0.32 & 0.20 & 0.19 & 0.34 & 0.24 & 0.41 & 0.59 \\ 
			GMEXICOB-MXX & 0.29 & 0.29 & 0.26 & 0.45 & 0.29 & 0.33 & 0.22 & 0.30 & 0.25 & 0.50 \\ 
			ASURB-GAPB & 0.24 & 0.17 & 0.16 & 0.15 & 0.21 & 0.33 & 0.20 & 0.33 & 0.32 & 0.32 \\ 
			ICHB-SIMECB & 0.34 &  & 0.44 & 0.34 & 0.11 & 0.37 &  & 0.37 & 0.25 &  \\ 
			\hline
		\end{tabular}
	\end{adjustbox}
\end{table}


\providecommand{\bysame}{\leavevmode\hbox to3em{\hrulefill}\thinspace}
\providecommand{\MR}{\relax\ifhmode\unskip\space\fi MR }
\providecommand{\MRhref}[2]{%
	\href{http://www.ams.org/mathscinet-getitem?mr=#1}{#2}
}
\providecommand{\href}[2]{#2}

\end{document}